\documentclass[sigconf]{acmart}

\usepackage{amsmath,amsfonts}
\usepackage{lettrine}
\usepackage{multirow}
\usepackage{multicol}
\usepackage{tablefootnote}
\usepackage{enumerate}
\usepackage{enumitem}
\usepackage{extarrows}
\usepackage{booktabs}
\usepackage{mathtools}
\usepackage{caption}
\usepackage{graphicx}
\usepackage{url} 
\usepackage{pifont}
\usepackage{marvosym}
\usepackage{xcolor}

\newtheorem{Lemma}{Lemma}

\usepackage{booktabs}   
\usepackage[table]{xcolor} 
\usepackage{amsmath}    
\usepackage{wrapfig}
\definecolor{baselinegray}{gray}{0.95}
\definecolor{improvementblue}{cmyk}{0.1, 0, 0, 0}
\definecolor{textgray}{gray}{0.55} 
\definecolor{deltablue}{cmyk}{1, 0, 0, 0} 
\definecolor{classcolor}{HTML}{FCE5CD}
\definecolor{vqacolor}{HTML}{D9E1F2}
\definecolor{retcolor}{HTML}{E2EFDA}
\definecolor{grdcolor}{HTML}{F2DCDB}
\definecolor{finalcolor}{HTML}{DEEBF6}
\definecolor{oodcolor}{HTML}{FFF2CC}
\AtBeginDocument{%
  }

\setcopyright{acmlicensed}
\copyrightyear{2018}
\acmYear{2018}
\acmDOI{XXXXXXX.XXXXXXX}
\begin{document}

\title{Leveraging Scene Context with Dual Networks for Sequential User Behavior Modeling}

\author{Xu Chen*, Yunmeng Shu*, Yuangang Pan, Jinsong Lan\\Xiaoyong Zhu, Shuai Xiao, Haojin Zhu, Ivor W. Tsang, Bo Zheng}
\email{{huaisong.cx,shuyunmeng.sym,jinsonglan.ljs,xiaoyong.z,shuai.xsh,bozheng}@alibaba-inc.com,
zhuhaojin@gmail.com,
{pan_yuangang,ivor_tsang}@cfar.a-star.edu.sg}
\affiliation{%
  \institution{Alibaba Group, China and A$^*$ STAR, Singapore}
  \country{}
}

\renewcommand{\shortauthors}{Xu Chen et al.}

\begin{abstract}
Modeling sequential user behaviors for future behavior prediction is crucial in improving user's information retrieval experience. Recent studies highlight the importance of incorporating contextual information to enhance prediction performance. One crucial but usually neglected contextual information is the scene feature which we define as sub-interfaces within an app, created by developers to provide specific functionalities, such as ``text2product search" and ``live" modules in e-commence apps. Different scenes exhibit distinct functionalities and usage habits, leading to significant distribution gap in user engagement across them. Popular sequential behavior models either ignore the scene feature or merely use it as attribute embeddings, which cannot effectively capture the dynamic interests and interplay between scenes and items when modeling user sequences. In this work, we propose a novel Dual Sequence Prediction networks (DSPnet) to effectively capture the dynamic interests and interplay between scenes and items for future behavior prediction. DSPnet consists of two parallel networks dedicated to learn users' dynamic interests over items and scenes, and a sequence feature enhancement module to capture the interplay for enhanced future behavior prediction. Further, we introduce a Conditional Contrastive Regularization (CCR) loss to capture the invariance of similar historical sequences. Theoretical analysis suggests that DSPnet is a principled way to learn the joint relationships between scene and item sequences. Extensive experiments are conducted on one public benchmark and two collected industrial datasets. The method has been deployed online in our system, bringing a 0.04 point increase in CTR, 0.78\% growth in deals, and 0.64\% rise in GMV.
The codes are available at this anonymous github: \textcolor{blue}{\url{https://anonymous.4open.science/r/DSPNet-ForPublish-2506/}}.
\end{abstract}




\maketitle


\section{Introduction}
Modern online information retrieval services, such as search and recommendation, have brought great changes and convenience for human's daily life. Correspondingly, users' sequential behaviors spread over a variety of apps and websites~\cite{9404857,9837877}. Modeling these behaviors for future behavior prediction has become an important issue in real-world applications.

Recent advances in modeling sequential user behaviors concentrate on three key areas: \textit{design of the encoding architecture}, \textit{formulation of the training objective} and \textit{utilization of the contextual information}. 
In \textit{design of the encoding architecture}, early works employ Markov models~\cite{rendle2010} to capture sequential patterns within historical behavior sequences. However, these models face limitations in their ability to represent complex and higher-order sequential dependencies. Consequently, researchers tend to investigate the more expressive recurrent neural networks (RNNs)~\cite{medsker2001recurrent,hidasi2016,Hidasi2018,10.1145/3109859.3109877} or self-attention mechanisms~\cite{attention_is_all,kang2018self,bert4rec}, to enhance sequential behavior modeling. 
Subsequently, researchers explored more advanced \textit{formulation of training objective}, beyond the conventional next-item prediction objective. They primarily designed various self-supervised training tasks to extract additional insights from sequences during training~\cite{bert4rec,10.1145/3459637.3481952,s3_rec,msdp,contracRec}.
Additionally, several studies have focused on \textit{utilization of the contextual information} such as item category~\cite{cocoRec}, behavior type~\cite{csan,chen2023survey} and time intervals~\cite{ye_2020}, as the contextual information notably influences user behaviors. 

\begin{figure*}[t]
\centering
\begin{minipage}[t]{0.21\textwidth}
\centering
\includegraphics[width=\textwidth]{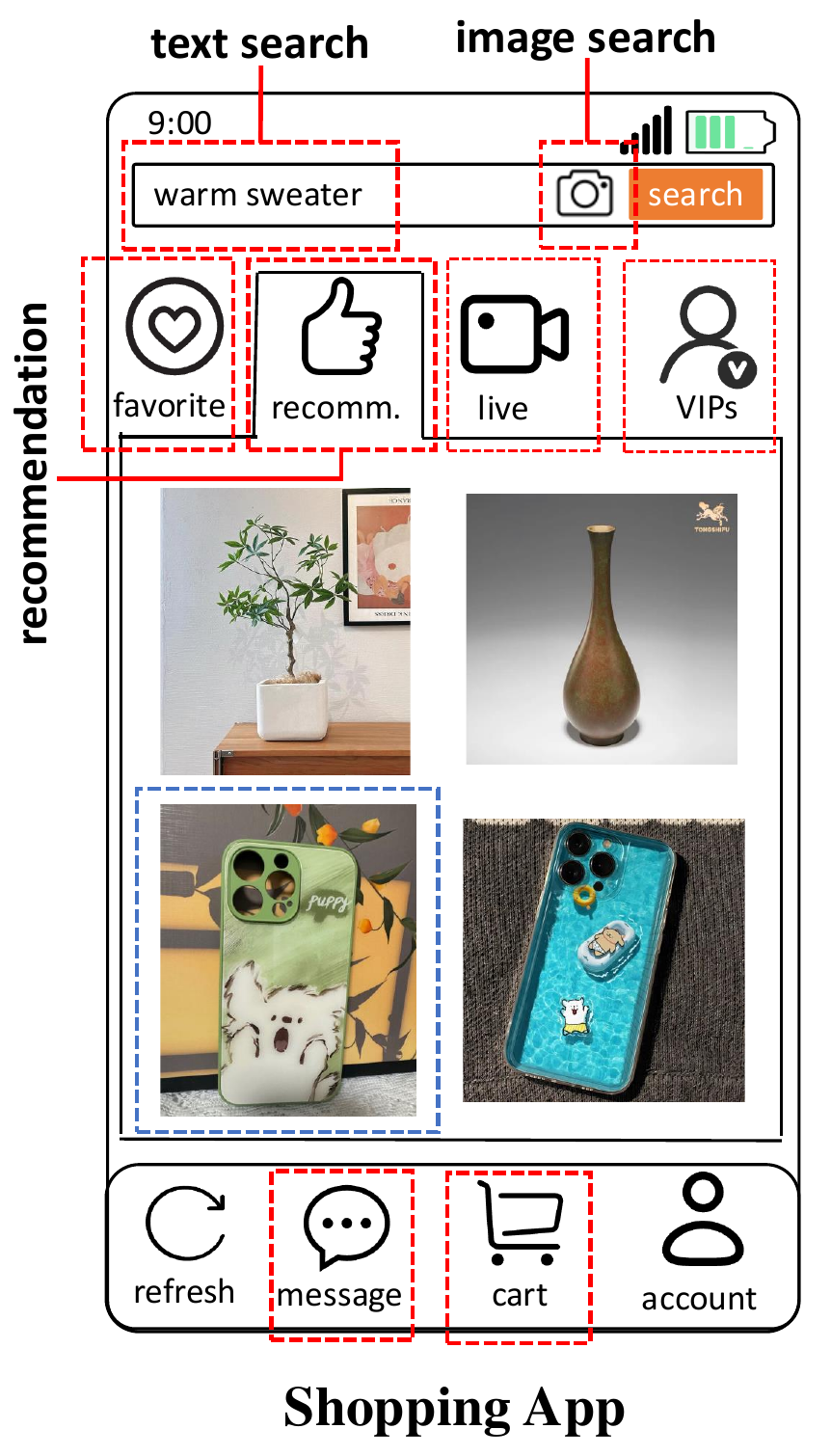}
\vspace{-18pt}
\caption*{{(a)}}
\end{minipage}
\hspace{10pt}
\begin{minipage}[t]{0.23\textwidth}
\centering
\includegraphics[width=\textwidth]{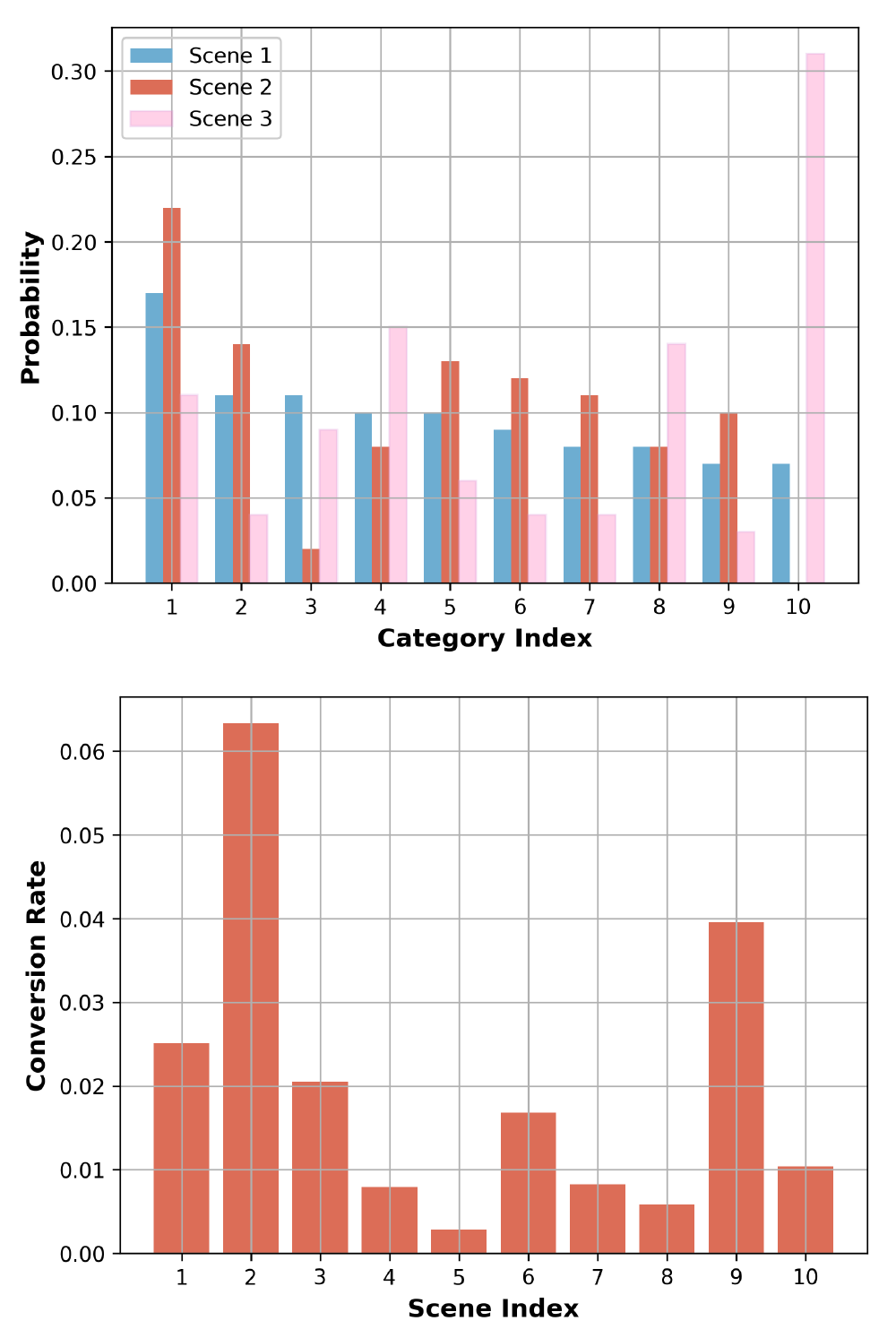}
\vspace{-18pt}
\caption*{{(b)}}
\end{minipage}
\hspace{5pt}
\begin{minipage}[t]{0.28\textwidth}
\centering
\includegraphics[width=\textwidth]{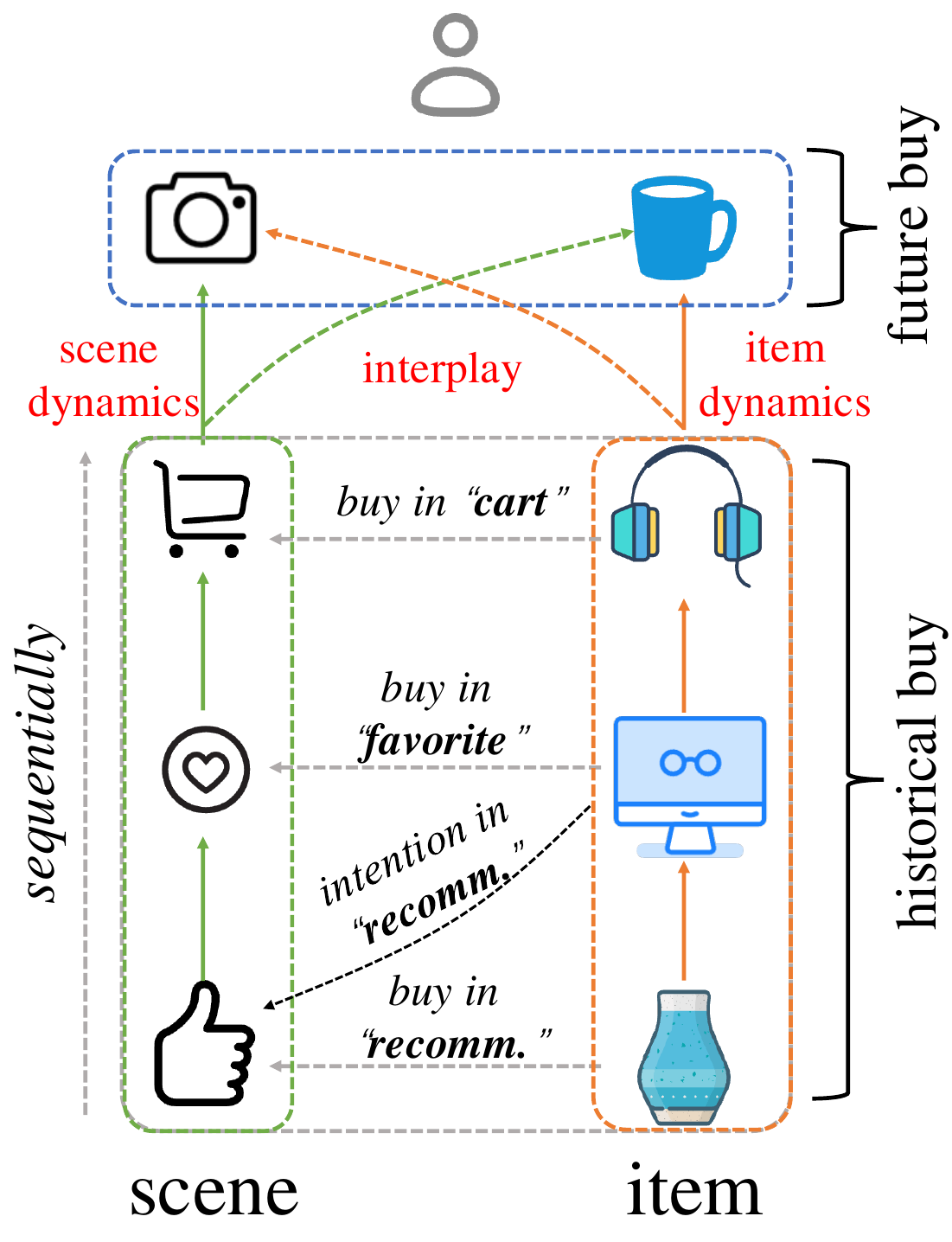}
\vspace{-20pt}
\caption*{{(c)}}
\end{minipage}
\vspace{-10pt}
\caption{(a) An example to show different scenes (\textit{i.e.} interfaces) within an App. The red dashed boxes represent distinct scenes, while the blue dashed boxes indicate individual items. (b) An example to show the distribution gap among scenes. The upper one indicates category distribution across scenes in our e-commence app. The lower one shows users' conversion rate in different scenes.
(c) shows a purchase sequence across different scenes. When buying a monitor, the user showed purchase intention in the ``recommendation" scene but made the purchase later in the "favorite" scene due to a sales promotion a few days afterward. This example highlights the intention misalignment issue of each scene-item behavior.}
\label{fig:intro_plot}
\vspace{-12pt}
\end{figure*}

Different from the above contextual feature, one crucial but usually neglected contextual factor influencing user behaviors is the scene feature, \textbf{which we define as sub-interfaces created by developers to encapsulate specific themes or functionalities within apps or websites.}
As shown in Figure~\ref{fig:intro_plot} (a),
different scenes, usually operated by different teams, have different themes and styles. For example, the shopping app encompasses scenes such as ``text search", ``recommendation" and ``live", facilitating functionalities like text-to-product search, product recommendations, and interactive live shopping experiences. 
These scenes represent different shopping types, leading to significant distribution gap in user engagements.
Figure~\ref{fig:intro_plot} (b) illustrates the category distribution and user conversion rate across different scenes, unveiling significant disparities in both item content and user engagement among scenes. This scene information is essential for delivering conditional information that influences behavior occurrences.
When a user enters a specific scene, it can reflect certain shopping interests and the interests greatly impact the items the user is likely to interact with.

\textit{Ignoring the scene feature would result in a large loss of information and introduce data bias in modeling sequential user behaviors.} 
Currently, there are limited studies addressing this problem because of limited data sources. \textit{There remains two challenges to effectively incorporate this scene feature in sequential behavior modeling.} 
1) \textit{One challenge is that how to better learn sequence dynamics by considering the intention misalignment issue of scene-item data.} To be specific, the misalignment issue means a user's interest and intention are initially reflected in one scene but the behavior are incorrectly collected in another scene. For example in Figure~\ref{fig:intro_plot} (c), in an e-shopping app, a user might see one monitor in the ``recommendation" interface with the intention to purchase it. However, instead of buying immediately, the user added it to the ``favorite" interface and bought from the ``favorite" interface later, because of upcoming sales promotions from sellers. 
2) \textit{The other challenge is that how to capture the interplay between scenes and items.} Although we may incorporate the scene feature as an input field of items by following~\cite{multi_sceneraio,carcaRec}, it could overlook the interplay between scene and item, which hinders their ability of mining scene feature in behavior prediction. As shown in Figure~\ref{fig:intro_plot} (c), the item and scene simultaneously occur in an interaction, and sequential scene dynamics and item dynamics have mutual effects in generating subsequent behaviors.
Further, the misalignment issue would bring obstacles when modeling the interplay between scenes and items.


To this end, we propose a novel Dual Sequence Prediction network (DSPnet) that effectively captures the interplay between sequential scenes and items, while being robust against the intention misalignment issue of scene-item data to predict future user behaviors. 
\textit{DSPnet comprises two parallel networks dedicated to predicting scene and item sequences, together with a sequence feature enhancement module to deliver the mutual effects across both sequences.} 
In particular, the scene sequence prediction network and item sequence prediction network encode their own dynamics from historical behaviors and maintain intention consistency in sequence level, which avoids the one-to-one scene-item intention misalignment in learning. Meanwhile, the sequence feature enhancement module enables one network's encoding features to be input into the other, allowing both prediction networks to capitalize on their interplay during the sequence learning process. 
We also demonstrate that the learning approach of DSPnet is theoretically equivalent to maximizing the joint log-likelihood of scenes and items, presenting a good way to model their relationships and sequential dynamics.
Given that sequential behaviors often exhibit randomness and noise, which can adversely affect the learning of sequence dynamics, we introduce Conditional Contrastive Regularization (CCR) loss to capture the representation invariance of similar historical sequences. 
Through learned conditional weights, CCR loss can adaptively promote similarity in sequence representations that undergo augmentation with different forces. We empirically demonstrate CCR loss highlights the relationships among contrasting samples, enhancing the model’s robustness in representation learning for real-world, skewed behaviors. 
We have conducted an online A/B test in our e-shopping search system, which shows that DSPnet could achieve an absolute 0.04 point CTR improvement, along with a relative 0.78\% deal growth and a relative 0.64\% Gross Merchandise Value (GMV) rise.
The contributions are summarized as follows:
\begin{itemize}
\vspace{-10pt}
    \item We focus on the feature of sub-interfaces in apps or websites for modeling sequential behaviors, referring to this as the scene feature. We propose a novel DSPnet that enhances sequential user behavior prediction with this scene feature. 
    \item We introduce CCR loss to enhance the model's representation learning by capturing the invariance of similar historical sequences. CCR uses learned conditional weights to more effectively promote representation performance in skewed user behaviors. 
    \item We conduct experiments on one public benchmark and two collected industrial datasets. Results show the impact of employing scene information in sequential behavior modeling and how our method outperforms state-of-the-art baselines.
\vspace{-6pt}
\end{itemize}

\section{Related Work}

\textbf{Design of the Encoding Architecture}: Early works investigate Markov chains~\cite{ching2006markov} to capture sequential dynamics within historical sequential behaviors. 
However, as the number of past actions increases, the state space grows exponentially, making it challenging to capture higher-order dependencies in real-world applications. Consequently, researchers explore more expressive neural sequence models like Recurrent Neural Networks (RNNs)~\cite{medsker2001recurrent,hidasi2016,Hidasi2018,10.1145/3109859.3109877}, Convolutional Neural Networks (CNNs)~\cite{caser}, Long Short-Term Memory Networks (LSTM)~\cite{graves2012long,duan2023long} and self-attention~\cite{attention_is_all,kang2018self,bert4rec} models to enhance sequential behavior modeling. 
For example,
Caser~\cite{caser} represents the recent item sequence in an ``image'' format and employs convolutional filters to capture local sequential patterns effectively.
\cite{hidasi2016} introduces a ranking loss function to classic RNNs, which makes it more viable for next item prediction. 
SASRec~\cite{kang2018self} and BERT4Rec~\cite{bert4rec} broadened the application of self-attention models to sequential behavior modeling. 
Some works~\cite{hu2024enhancing,li2024calrec,zheng2024harnessing,liao2024llara} focus on leveraging large language models (LLMs) for sequential recommendation, including aligning sequential recommendation systems with LLMs, summarizing user preferences. Others~\cite{ma2024plug,yang2024generate,wang2024conditional} investigate the application of diffusion models in sequential recommendation, aiming to better capture the evolution of user preferences over time. LLMs and diffusion based sequential recommendation models are beyond the scope of this paper, so we do not elaborate more here.

\textbf{Formulation of the Training Objective}: 
Several studies concentrate on forecasting item lists over specific time periods or behavioral distributions instead of the next individual items. 
SUMN~\cite{gu2021exploiting} hypothesizes that future behavioral distributions should align with past distributions. It learns sequence representations by maximizing the Kullback-Leibler divergence between item occurrence distributions from a previous period and those of the future. MSDP~\cite{msdp} uses a multi-scale approach to optimize predictions for the next period by considering item lists across different timeframes. Constructing self-supervised learning tasks to facilitate the prediction of sequential user behaviors has also gained considerable attention.
\cite{yao2021self} introduces a self-supervised learning framework that leverages feature correlations among items by using mask and dropout sequences. S3-Rec~\cite{s3_rec} implements four self-supervised tasks using raw features from items.
CL4SRec~\cite{DuoRec} explores the contrastive signals derived from augmented historical sequences through contrastive learning. ContraRec~\cite{wang2023sequential} achieves remarkable performance by constructing contrastive sequences using random mask and reorder augmentation techniques.

While contrastive learning improves sequential behavior modeling, it often ignores the varying importance of positive and negative samples. Our CCR loss learns conditional weights for these samples, capturing their unique contributions and enhancing the model’s robustness in learning sequence dynamics.


\textbf{Utilization of the Contextual Information}: 
In user behavior sequences, there are various contextual factors linked to each action, such as types of user behavior (e.g., clicks, purchases, additions to favorites)~\cite{mkm_sr,dupn,10.1145/3581783.3611723,10.1109/TKDE.2022.3175094,10.1145/3539597.3570386,chen2023survey}, product category~\cite{cocoRec} and other multiple item attributes~\cite{carcaRec}.
DUPN~\cite{dupn} incorporates multiple kinds of behavior types to construct multi-task learning for more effective personalization. \cite{10.1145/3539597.3570386} designed a multi-behavior learning module to extract users' personalized information for user-embedding enhancement. \textbf{These works study different behavior types, which is orthogonal to ours focusing on the modeling of defined scene feature.}
MKM-SR~\cite{mkm_sr} points out that a user's sequence behaviors could have some micro-behaviors that reflect fine-grained and deep understanding of the user's preference. 
The micro-behaviors are identified by specific behavior activities (\textit{e.g.} reading comments, adding to cart), and embedded individually to augment the original sequence prediction. 
CoCoRec~\cite{cocoRec} leverages item category to organize a user's own past actions and further employs self-attention to capture in-category transition patterns. These transition patterns are used to find similar users, enhancing collaborative learning. CARCA~\cite{carcaRec} incorporates both the attributes of interacted items and contextual data of user interactions by employing combined sequences as input for multi-head attention blocks. 
\textbf{Some other works employ different scene definitions from ours.} \cite{chen2021scene} explored adaptive sequential recommendation systems (RS) across different domains. \cite{Wang2021SceneRecSG} defined the scene as a collection of predefined item categories. \cite{10.1145/3640457.3688135} investigated the usage of LLMs for real-time problem in sequential RS. \cite{10.1145/3640457.3688115} defined scenes as 200 predefined topics, such as ``weekend spring outing'', ``afternoon tea'' and ``KFC crazy Thursday''.

\textbf{Above works studied different kinds of contextual information from our defined scene feature.} Integrating our scene feature into sequential behavior modeling constitutes an important and new problem derived from real-world applications. Although we may leverage the scene as an additional attribute of items by following~\cite{multi_sceneraio,carcaRec}, it would ignore the 
intention misalignment issue and fail to well capture the interplay between scenes and items.

\begin{figure*}[t]
\centering
\includegraphics[width=15.0cm]{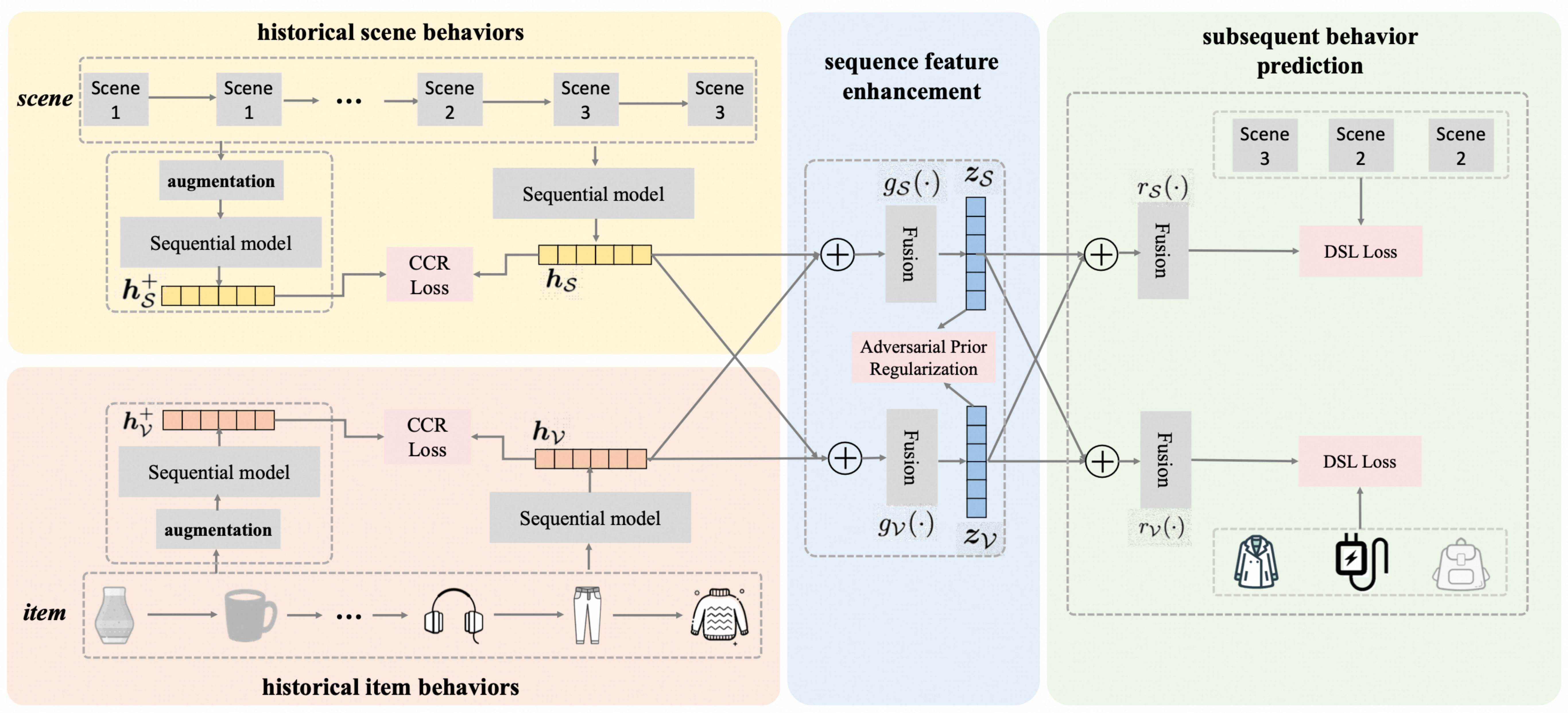}
\vspace{-8pt}
\caption{
The architecture of DSPnet. Its dual sequence learning models the interplay between scene and item sequences while capturing the sequence dynamics against user intention misalignment issue of scene-item data. CCR loss learns representation invariance with different forces on different samples. $\oplus$ means concatenation.}
\label{figure:model_architecture}
\vspace{-15pt}
\end{figure*}
\section{Dual Sequence Prediction Network}
\subsection{Overview}
In sequential user behavior prediction, we aim to predict a user's future behaviors based on historical behaviors. Given a historical behavior sequence $\mathcal{T}$ from user $\boldsymbol{u}$, it is defined as:
\begin{align}
    \mathcal{T}=\{(\boldsymbol{v}_1,\boldsymbol{s}_1),(\boldsymbol{v}_2,\boldsymbol{s}_2),...,(\boldsymbol{v}_j,\boldsymbol{s}_j),...,(\boldsymbol{v}_{|\mathcal{T}|},\boldsymbol{s}_{|\mathcal{T}|}) \},
\end{align}
where $\boldsymbol{v}_j\in \boldsymbol{V} = \{\boldsymbol{v}_1,...,\boldsymbol{v}_{N_{v}}\}$ and $\boldsymbol{s}_j\in \boldsymbol{S} = \{\boldsymbol{s}_1,...,\boldsymbol{s}_{N_{s}} \}$ denote one interacted item and the corresponding scene, respectively.
$\boldsymbol{V}$ denotes the whole item set with size $N_{v}$ and $\boldsymbol{S}$ denotes the whole scene set with size $N_{s}$. $|\mathcal{T}|$ is the number of historical interactions. The historical behaviors actually contains two coupling sequences, \textit{i.e.} the item sequence $\mathcal{V}=\{\boldsymbol{v}_1,\boldsymbol{v}_2,...,\boldsymbol{v}_{|\mathcal{T}|}\}$ and the scene sequence $\mathcal{S}=\{\boldsymbol{s}_1,\boldsymbol{s}_2,...,\boldsymbol{s}_{|\mathcal{T}|}\}$.
We aim to predict future behaviors based on the historical behaviors.
\textbf{To better study the issue, we only consider one-type behavior which means the sequence includes only one-type behavior, \textit{e.g.} ``buy''.}

DSPnet consists of two main and original components: dual sequence learning and conditional contrastive regularization (CCR) loss. Both are designed to tackle important challenges. The first one is proposed to effectively encode sequential dynamics against user intention misalignment issue of scene-item behaviors and deliver the dynamics to both scene and item sides for predicting behaviors. The second one aims to learn representation invariance and dynamically enhance the similarity between representations of similar sequences, thereby improving the model's robustness against random and skewed user behaviors. The model architecture is shown in Figure~\ref{figure:model_architecture}, details are in the following parts.


\subsection{Dual Sequence Learning}
\textbf{Dual Sequence Dynamics}: 
Online user behaviors occur under different contexts chronologically, reflecting users' dynamic interests over time. 
The defined scene feature largely influences how users interact with items, while interactions taken on particular items can also influence users' decisions in following scenes. Therefore, the sequential dynamics of both items and scenes play a vital role in predicting subsequent user behaviors. \textit{Considering the intention misalignment issue of scene-item data caused by many factors such as feedback delay and seller promotions, we encode the dynamics in sequence-level correspondence, rather than maintaining strict one-to-one scene-item correspondence.} Given the historical scene sequence $\mathcal{S}$ and item sequence $\mathcal{V}$, we employ sequential models to capture the dynamics. Let $f_{\mathcal{S}}$ and $f_{\mathcal{V}}$ be the sequential models of historical scenes and items, we have:
\begin{align}
    \boldsymbol{h}_{\mathcal{S}} = f_{\mathcal{S}}(\mathcal{S}),~\boldsymbol{h}_{\mathcal{V}} = f_{\mathcal{V}}(\mathcal{V}),
\end{align}
where $\boldsymbol{h}_{\mathcal{S}}$ and $\boldsymbol{h}_{\mathcal{V}}$ mean the encoded latent representations of $\mathcal{S}$ and $\mathcal{V}$, respectively. The choice of sequential model is flexible (such as RNN, LSTM and transformer) and we employ the powerful transformer model following recent works~\cite{bert4rec,msdp,contracRec}.

\textbf{Sequence Feature Enhancement}: As we previously discussed, both the item and scene mutually influence subsequent behavior occurrence. This means that when predicting future interacted items, we cannot solely depend on historical item interactions. Similarly, if we predict the subsequent interacted scenes, we cannot only employ past scene interactions. Either the item sequence or the scene sequence serves as an enhanced information for the other predictions. 
Denoting $\boldsymbol{z}_{\mathcal{S}}$ and $\boldsymbol{z}_{\mathcal{V}}$ as the feature enhanced representation of scene sequence and item sequence, we have:
\begin{align}
\label{eq:sequential_encoder}
    \boldsymbol{z}_{\mathcal{S}} = g_{\mathcal{S}}(\boldsymbol{h}_{\mathcal{S}}\oplus  \boldsymbol{h}_{\mathcal{V}}),~~\boldsymbol{z}_{\mathcal{V}} = g_{\mathcal{V}}(\boldsymbol{h}_{\mathcal{S}}\oplus  \boldsymbol{h}_{\mathcal{V}}),
\end{align}
where $g_{\mathcal{S}}(\cdot)$ and $g_{\mathcal{V}}(\cdot)$ denote the fusion MLP layers.
Both $\boldsymbol{z}_{\mathcal{S}}$ and $\boldsymbol{z}_{\mathcal{V}}$ can be considered as the user representation $\boldsymbol{z_{u}}$ that encodes mutual effects and dynamics from historical behaviors. 
\textit{We maintain these two enhanced representations here to provide diverse aspects of the user interests and better facilitate subsequent item and scene prediction tasks.}

\textbf{Subsequent Behavior Prediction}: When conducting subsequent behavior prediction, we incorporate both enhanced representations $\boldsymbol{z}_{\mathcal{S}}$ and $\boldsymbol{z}_{\mathcal{V}}$ for behavior prediction:
\begin{align}
\label{eq:feature_selection_prediction}
    \boldsymbol{o}_{\mathcal{S}}=r_{\mathcal{S}}(\boldsymbol{z}_{\mathcal{S}}\oplus \boldsymbol{z}_{\mathcal{V}}),~~\boldsymbol{o}_{\mathcal{V}}=r_{\mathcal{V}}(\boldsymbol{z}_{\mathcal{S}}\oplus \boldsymbol{z}_{\mathcal{V}}),
\end{align}
where $\boldsymbol{o}_{\mathcal{S}}$ and $\boldsymbol{o}_{\mathcal{V}}$ are outputs of the feature selection functions.
Additionally, since the next user behavior may be stochastic and noisy, we choose to predict the subsequent behaviors over a period of time by following~\cite{msdp}. Here, we denote the candidate scene set and candidate item set for prediction as $\boldsymbol{V}^{\text{cand}}\subseteq \boldsymbol{V}$ and $\boldsymbol{S}^{\text{cand}}\subseteq \boldsymbol{S}$, respectively. The ground-truth label of subsequent scene behavior is given by $\boldsymbol{y^s}\in \{0,1\}^{K^s}$ and the subsequent item behavior is given by $\boldsymbol{y^v}\in \{0,1\}^{K^v}$, where $K^s$ and $K^v$ are the size of candidate scene or item set. 1 means positive feedback while 0 is negative feedback. Then the prediction objective functions on future scenes and items are formulated as follows:
\begin{subequations}
\label{eq:feat_enhance}
\begin{align}
     \mathcal{L}_{\text{DSL}}^{\mathcal{S}} &= -\frac{1}{K}\sum_{k=1}^K
    [\boldsymbol{y^s}_{k} \log (\hat{\boldsymbol{y^s}_{k}}) + (1 - \boldsymbol{y^s}_{k}) \log (1 - \hat{\boldsymbol{y^s}_{k}})], \\
   \mathcal{L}_{\text{DSL}}^{\mathcal{V}} &= -\frac{1}{K}\sum_{k=1}^K
    [\boldsymbol{y^v}_{k} \log (\hat{\boldsymbol{y^v}_{k}}) + (1 - \boldsymbol{y^v}_{k}) \log (1 - \hat{\boldsymbol{y^v}_{k}})],
\end{align}
\end{subequations}
where $\hat{\boldsymbol{y^s}_{k}}=\sigma( \boldsymbol{o}_{\mathcal{S}} \cdot \boldsymbol{e^s}_{k})$ and $\hat{\boldsymbol{y^v}_{k}}=\sigma( \boldsymbol{o}_{\mathcal{V}} \cdot \boldsymbol{e^v}_{k})$ indicate the prediction probability of $k$-th scene and item in candidate sets, $\boldsymbol{e^s}_{k}$ and $\boldsymbol{e^v}_{k}$ are the latent representations of $k$-th candidate scene and item, respectively. $\sigma(\cdot)$ is the sigmoid function and $\cdot$ means inner-product operation.

\textbf{Adversarial Prior Regularization}: Since user behaviors usually face severe data sparsity problem and user representations may overfit to samples, we impose adversarial prior regularization on learned user representations~\cite{aae,xu_towards2023}. This approach is more advantageous than Kullback-Leibler (KL) divergence regularization because of its flexibility and easy implementation. 

Let $D_{\mathcal{S}}$ and $D_{\mathcal{V}}$ be the discriminator of $\boldsymbol{z}_{\mathcal{S}}$ and $\boldsymbol{z}_{\mathcal{V}}$ respectively, the adversarial learning based prior regularization is written as:
\begin{small}
\begin{align}
\label{eq:adv_reg}    &\min_{g_{\mathcal{S}},g_{\mathcal{V}},f_{\mathcal{S}},f_{\mathcal{V}}}~\max_{D_{\mathcal{S}},D_{\mathcal{V}}} \mathcal{L}_{\text{APR}} = \nonumber \\
& \mathbb{E}_{\boldsymbol{z}_{\mathcal{S}}\sim p(\boldsymbol{z}_{\mathcal{S}})}[\log D_{\mathcal{S}}(\boldsymbol{z}_{\mathcal{S}})] +  \mathbb{E}_{\boldsymbol{z}_{\mathcal{S}}\sim g_{\mathcal{S}(\cdot)}}[\log (1 - D_{\mathcal{S}}(\boldsymbol{z}_{\mathcal{S}}))] \nonumber \\
& +\mathbb{E}_{\boldsymbol{z}_{\mathcal{V}}\sim p(\boldsymbol{z}_{\mathcal{V}})}[\log D_{\mathcal{V}}(\boldsymbol{z}_{\mathcal{V}})] +  \mathbb{E}_{\boldsymbol{z}_{\mathcal{V}}\sim g_{\mathcal{V}(\cdot)}}[\log (1 - D_{\mathcal{V}}(\boldsymbol{z}_{\mathcal{V}}))],
\end{align}
\end{small}

where $p(\boldsymbol{z}_{\mathcal{S}})$ and $p(\boldsymbol{z}_{\mathcal{V}})$ are the prior distributions. DSPnet has a behavior prediction task that matches complex behavior distributions, largely preventing mode collapse in adversarial learning.


\textbf{Theoretical Analysis}: We also reveal the theoretical analysis of this dual sequence learning mechanism on capturing the interplay between scene and item sequences. 
\begin{Lemma}
\label{lemma:1}
\textit{Without specifying the sequential encoder architecture and prediction objective function, minimizing the dual sequence learning scheme is equivalent to maximizing the following \textit{evidence lower bound} of the joint log-likelihoods of observed item and scene sequential behaviors:}
\begin{small}
\begin{align}
\label{eq:lemma1}
    &\max_{\theta_1,\theta_2,\phi_1,\phi_2} \mathcal{L}_{\text{ELBO}} = \nonumber \\
    &\mathbb{E}_{q_{\phi_1}(\boldsymbol{z}_{\mathcal{V}}|\mathcal{V},\mathcal{S})q_{\phi_2}(\boldsymbol{z}_{\mathcal{S}}|\mathcal{V},\mathcal{S})}[\log p_{\theta_1}(\boldsymbol{v}|\boldsymbol{z}_{\mathcal{V}},\boldsymbol{z}_{\mathcal{S}})p_{\theta_2}(\boldsymbol{s}|\boldsymbol{z}_{\mathcal{V}},\boldsymbol{z}_{\mathcal{S}})] \nonumber \\
    & - D_{KL}[q_{\phi_1}(\boldsymbol{z}_{\mathcal{V}}|\mathcal{V},\mathcal{S})||p(\boldsymbol{z}_{\mathcal{V}})] - D_{KL}[q_{\phi_2}(\boldsymbol{z}_{\mathcal{S}}|\mathcal{V},\mathcal{S})||p(\boldsymbol{z}_{\mathcal{S}})],
\end{align}  
\end{small}

where $\boldsymbol{v}$ and $\boldsymbol{s}$ denote the observed item and scene, respectively. $\mathcal{V}$ and $\mathcal{S}$ indicate the historical sequential items and scenes before $\boldsymbol{v}$ and $\boldsymbol{s}$, respectively. $\boldsymbol{z}_\mathcal{V}$ and $\boldsymbol{z}_\mathcal{S}$ are the encoded representations from historical behaviors. $D_{KL}$ is the KL Divergence that conducts prior regularization.
\end{Lemma}
Detailed derivation is given in Appendix~\ref{appendix:joint_log_likelihood}. The meaning of Eq.~\ref{eq:lemma1} is equivalent to our model design in Figure~\ref{figure:model_architecture}. To be specific, $q_{\phi_1}(\boldsymbol{z}_{\mathcal{V}}|\mathcal{V},\mathcal{S})$ and $q_{\phi_2}(\boldsymbol{z}_{\mathcal{S}}|\mathcal{V},\mathcal{S})$ are the posteriors that encode information from $\mathcal{V},\mathcal{S}$ into $\boldsymbol{z}_{\mathcal{V}},\boldsymbol{z}_{\mathcal{S}}$. 
$p_{\theta_1}(\boldsymbol{v}|\boldsymbol{z}_{\mathcal{V}},\boldsymbol{z}_{\mathcal{S}})$ and $p_{\theta_2}(\boldsymbol{s}|\boldsymbol{z}_{\mathcal{V}},\boldsymbol{z}_{\mathcal{S}})$ indicate that predicting the future items or scenes both should be dependent on the historical behaviors, like our design in Eq.~\ref{eq:feature_selection_prediction}. 
The last two KL divergence terms correspond to our adversarial learning based prior regularization in Eq.~\ref{eq:adv_reg}.

\textbf{\underline{\emph{Remark}}}: \textbf{Modeling the joint log-likelihood of scene and item sequences is a principled way to capture their interplay and dynamics, which is usually overlooked in~\cite{bert4rec,msdp,chen2019cascading,contracRec}.} Although some works~\cite{multi_sceneraio,carcaRec} can incorporate the scene information as an additional attribute embedding for items, they are sensitive to the intention misalignment issue of scene-item pairs and fail to better capture such interplay, and we verify this through following experiments. In this regard, dual sequence learning can empower our DSPnet to learn more comprehensive representations of historical behaviors for improved behavior predictions.

\subsection{Sequential Contrastive Learning}
\label{sec:CCR}
Unlike human language, sequential user behaviors are often random and noisy. To better capture the dynamics of behavior sequences, we use sequential contrastive learning, containing two components: sequence augmentation and conditional contrastive regularization.  


\textbf{Sequence Augmentation}: The sequence augmentation must not alter the user's intention in input sequence. Drawing insights from recent studies~\cite{bert4rec,contracRec}, we utilize masking and reordering to perform sequence augmentation.
The masking augmentation involves randomly masking a percentage of elements from the input sequence. Reorder augmentation consists of two steps: first, we randomly select a size that ranges from 2 up to the length of the sequence. Then, we uniformly choose a continuous subsequence of this size and shuffle its elements, while the elements outside of this subsequence retain their original order.
Let $\mathcal{A}(\cdot)$ represent a function that applies augmentation to the original sequence. We can express the augmented historical scene and item sequence as $\mathcal{S}^{+}=\mathcal{A}(\mathcal{S})$ and $\mathcal{V}^{+}=\mathcal{A}(\mathcal{V})$, respectively. These augmented samples offer valuable signals for learning representation invariance.

\textbf{Conditional Contrastive Regularization}: 
To learn the invariance of behavior sequences, we aim to maximize the similarity between original and augmented sequences while minimizing the similarity of sampled dissimilar sequences. Additionally, we introduce two conditional weights to reflect the differing contributions of augmented and sampled dissimilar sequences in optimization.

Let $\boldsymbol{h}_{\mathcal{V}}^{+}$ be the representation of augmented item sequence $\mathcal{V}^{+}$ and $\boldsymbol{h}_{\mathcal{V}}^{-}$ be the representation of sampled dissimilar item sequence, our contrastive loss with two conditional weights is formulated as:
{\small
\begin{align}
    \mathcal{L}_{\text{CCR}}^{\mathcal{V}} =  &- \mathbb{E}_{\boldsymbol{h}_{\mathcal{V}}} [\sum_{i=1}^{N_+} 
    \underbrace{\frac{e^{-s(\boldsymbol{h}_{\mathcal{V}}, \boldsymbol{h}_{\mathcal{V},i}^+)}}{\sum_{i}^{N_+} e^{-s(\boldsymbol{h}_{\mathcal{V}}, \boldsymbol{h}_{\mathcal{V},i}^+)}} }_{\text{conditional weights:}~\boldsymbol{w}_{\mathcal{V},i}^{+}} s(\boldsymbol{h}_{\mathcal{V}}, \boldsymbol{h}_{\mathcal{V},i}^+) ] \nonumber \\
    &+  \mathbb{E}_{\boldsymbol{h}_{\mathcal{V}}} [ \sum_{j=1}^{N_-} 
    \underbrace{\frac{e^{s(\boldsymbol{h}_{\mathcal{V}}, \boldsymbol{h}_{\mathcal{V},j}^-)}}{\sum_{j}^{N_-} e^{s(\boldsymbol{h}_{\mathcal{V}}, \boldsymbol{h}_{\mathcal{V},j}^-)}}}_{\text{conditional weights:}~\boldsymbol{w}_{\mathcal{V},j}^{-}} s(\boldsymbol{h}_{\mathcal{V}}, \boldsymbol{h}_{\mathcal{V},j}^-) ],
\end{align}
}
where $s(\boldsymbol{h}_{\mathcal{V}},\boldsymbol{h}_{\mathcal{V}}^+)=\boldsymbol{h}_{\mathcal{V}}^T\boldsymbol{h}_{\mathcal{V}}^+/\tau^+$ and $s(\boldsymbol{h}_{\mathcal{V}},\boldsymbol{h}_{\mathcal{V}}^-)=\boldsymbol{h}_{\mathcal{V}}^T\boldsymbol{h}_{\mathcal{V}}^-/\tau^-$
calculate the similarity between two vectors. $\tau^+$ and $\tau^-$ are two temperature hyper-parameters. $N_{+}$ and $N_{-}$ indicate the number of augmented sequences (\textit{i.e.} positive samples) and dissimilar sequences (\textit{i.e.} negative samples). $\boldsymbol{w}_{\mathcal{V},i}^{+}$ and $\boldsymbol{w}_{\mathcal{V},j}^{-}$ are the conditional weights which are designed to mine hard samples to perform more effective optimization. The conditional contrastive loss of scene sequence $\mathcal{L}_{\text{CCR}}^{\mathcal{S}}$ can be written in similar formula with $(\boldsymbol{h}_{\mathcal{S}},\boldsymbol{h}_{\mathcal{S}}^{+},\boldsymbol{h}_{\mathcal{S}}^{-})$ as input. 

\textbf{\underline{\emph{Remark}}}: 
Given the original sequence, different augmented sequences could have different contributions in optimization. \textbf{When sampling negative samples for skewed data distributions, such as the pronounced long-tailed patterns in user behavior data, the relationships among negatives may be largely different from uniform distribution.} Therefore, it is vital to optimize the contrastive signals with conditional weights, unlike the uniform weights in conventional contrastive loss~\cite{align_uniform,simclr,contracRec}.

\subsection{Overall Objective Function}
To sum up, we can write the whole training objective function of DSPnet as follows:
\begin{align}
    \mathcal{L}_{\text{DSPnet}} = \mathcal{L}_{\text{DSL}}^{\mathcal{V}} +\lambda*\mathcal{L}_{\text{DSL}}^{\mathcal{S}} + \alpha * \mathcal{L}_{\text{APR}} + \beta * \mathcal{L}_{\text{CCR}},
\end{align}
where $\mathcal{L}_{\text{CCR}}=\mathcal{L}_{\text{CCR}}^{\mathcal{S}}+\mathcal{L}_{\text{CCR}}^{\mathcal{V}}$. The $\lambda$, $\alpha$ and $\beta$ are hyper-parameters to weight the importance of loss terms. \textit{We usually care more on future item prediction in practice, so we take $\mathcal{L}_{\text{DSL}}^{\mathcal{V}}$ to be the main part and set $\lambda$ on $\mathcal{L}_{\text{DSL}}^{\mathcal{S}}$ here.} 
To sum up, DSPnet offers an efficient and principled approach for modeling the interplay of sequential scene and item behaviors, while being robust against intention misalignment issue of scene-item data. \textbf{The model complexity analysis is given in Appendix~\ref{appen:complexity}.}


\section{Experiments and Analysis}
\subsection{Experiment Setup}
\label{sec:experiment_setup}

\begin{table}[]
\centering
\caption{The statistics of datasets.}
\vspace{-6pt}
\label{table:appendix_dataset}
\renewcommand{\arraystretch}{1.1}
 \setlength{\tabcolsep}{0.8mm}{ 
  \scalebox{0.85}{
\begin{tabular}{cccccc}
\hline
Dataset         & \#sequences & \#items    & \#scenes & \#avg. length & \#density \\ \hline
Outbrain        & 46,676      & 238,653    & 3,508     & 2.36          & 9.89e-4\% \\ \hline
AllScenePay-1m  & 1,000,000   & 7,871,700  & 330      & 25.35         & 3.22e-4\% \\ \hline
AllScenePay-10m & 10,000,000  & 32,766,762 & 801      & 25.33         & 7.70e-5\% \\ \hline
\end{tabular}
}}
\vspace{-12pt}
\end{table}
\textbf{Datasets}: 
We conduct experiments on three datasets, one of which is a public benchmark, while the other two are collected from our e-commence app. The public Ourbrain\footnote{\url{https://www.kaggle.com/competitions/outbrain-click-prediction/overview}} focuses on news recommendation and contains users' chronological views on documents. \textbf{For this dataset, we utilize the view sequence} and feature fields ``uuid", ``document\_id", ``timestamp", and ``source\_id". Here, ``uuid" identifies the user, while ``timestamp" records when an interaction occurred. The ``document\_id" serves as the item id, and ``source\_id", linked to the publisher's website, indicates the scene information. We take behaviors before 1975-10-01\footnote{This time is transformed from ``timestamp", without adding the time offset.} as historical behaviors and those following as prediction behaviors. We filtered sequences whose number of historical or future actions is less than 1. This dataset is split as train/val/test set with common 8/1/1 setting.


The two industrial datasets are named as AllScenePay-1m and AllScenePay-10m, which contain 1 million and 10 million \textbf{user purchase sequences}, respectively. In particular, we collected user purchase behaviors ranges from 2024-07-01 to 2024-08-07. We take behaviors from 2024-07-01 to 2024-07-31 as historical behaviors and those from 2024-08-01 to 2024-08-07 as prediction behaviors. We filtered out sequences with fewer than 3 historical or prediction behaviors, and formed the dataset AllScenePay.
Then we randomly sampled 1 million and 10 million user sequences from AllScenePay to construct those two experimental datasets. To conduct fast evaluation, we randomly select 10\% sequences from those two datasets as the val and test set, and the rest are taken as the train set. \textbf{More details including data statistics are given in Table~\ref{table:appendix_dataset}.}

\textbf{Baselines}: We make performance comparison with recent strong and popular methods, including the aspect of \textit{encoding architecture design}, \textit{training objective formulation} and \textit{contextual information utilization}. BERT4Rec~\cite{bert4rec} introduces bidirectional self-attention to sequential behavior modeling. MSDP~\cite{msdp} introduces a multi-scale stochastic distribution prediction as the training objective. ContracRec~\cite{contracRec} introduces a context-context contrastive loss to make similar sequences learn similar representations. Further, we introduce SceneCTC and SceneContraRec as the baselines incorporating scene information as embeddings like CARCA~\cite{carcaRec}. 
SceneCTC employs context-target contrastive loss from~\cite{contracRec}. SceneContraRec extends the input of ContraRec with scene feature. \textbf{We also add DSPnet-- that replaces our dual sequence encoder with one-to-one correspondence encoding in recent sequential RS to study the effectiveness of our dual sequence learning.}

\textbf{Parameter Settings}: 
The dimension of item embeddings and scene embeddings is set as 256 for all models on Outbrain. Since the number of items is too large on our industrial datasets, we set the dimension of item embeddings and scene embeddings as 16 and 4 on AllScenePay-1m and AllScenePay-10m for all models to save computation memory. 
We use one GPU for training on Outbrain and the batch size is 32.
While 8 GPUs are used on the two industrial datasets and the batch size on each GPU is 32.
We use the validation performance as early stop condition and the max training epoch is 100. Hyper-parameters of baselines are set according to their papers or searched on our datasets. 
In DSPnet, we employ the transformer in~\cite{bert4rec} as our sequential model and the transformer layer is 2. The number of MLP layers in $g_{\mathcal{S}}(\cdot)$ and $g_{\mathcal{V}}(\cdot)$ equals 2. 
The number of positive samples in CCR is 2, and that of negative samples is dependent on the batch size because we use the popular intra-batch sampling to sample negatives.
The temperature parameters are set as $\tau^+=1.0$ and $\tau^-=0.07$. 
Meanwhile, since the dataset size of Outbrain is small, we set $r_{\mathcal{S}}(\cdot)=r_{\mathcal{V}}(\cdot)$ as one linear MLP. 
We set $\lambda=1.0$, $\alpha=2\times 10^{-7}$ and $\beta=5\times 10^{-6}$ on Outbrain, while $\lambda=0.2$, $\alpha=10^{-9}$ and $\beta=10^{-7}$ on two industrial datasets\footnote{The value of $\mathcal{L}_{DSL}$ is quite small due to large number of negative samples. We set $\alpha$ and $\beta$ to a small scale to ensure they do not dominate $\mathcal{L}_{DSL}$.}. The prior distribution is standard Gaussian distribution.
\textbf{The study of other prior distributions is provided in Appendix~\ref{sec:appendix_prior}.}

\begin{table*}[]
\small
\centering
\caption{Performance comparison on next item prediction. R@$k$ and N@$k$ are Recall@$k$ and NDCG@$k$. We use ``w/o" to denote DSPnet without a particular part. The best results are bolded and \textit{the most competitive public baselines} are underlined.}
\vspace{-6pt}
\label{table:next_item_prediction}
\renewcommand{\arraystretch}{0.95}
 \setlength{\tabcolsep}{0.5mm}{ 
  \scalebox{0.9}{
\begin{tabular}{c|cccc|cccc|cccc}
\hline
Dataset               & \multicolumn{4}{c|}{Outbrain}                                         & \multicolumn{4}{c|}{AllScenePay-1m} & \multicolumn{4}{c}{AllScenePay-10m} \\ \hline
Method                & R@5             & N@5             & R@10            & N@10            & R@5    & N@5    & R@10    & N@10    & R@5    & N@5    & R@10    & N@10    \\ \hline
BERT4Rec              & 0.0943          & 0.0676          & 0.1384          & 0.0819          & {\small OOM}       &  {\small OOM}      &  {\small OOM}       &  {\small OOM}       & {\small OOM}       & {\small OOM}       & {\small OOM}        & {\small OOM}         \\
MSDP                                                                  & 0.2703                                                           & 0.2181                                                           & 0.2994                                                           & 0.2275                                                           & 0.0006  & 0.0004  & 0.0010  & 0.0005 &    0.0005  & 0.0003  & 0.0011 & 0.0005     \\
ContraRec                                                             & 0.3619                                                           & 0.2468                                                           & 0.4701                                                           & 0.2820                                                           & 0.0753  & 0.0533  & 0.1010  & 0.0616 & 0.1414       &  0.1026      & 0.1925        & 0.1191        \\
SceneCTC                                                                       & 0.4811                                                           & 0.4068                                                 & 0.5232                                                           & 0.4205                                                 & 0.0735                                                          & 0.0517                                                          & 0.1023                                                          & 0.0610                                                          & \underline{0.1459}  & 0.1027  & 0.1974 & 0.1193 \\
SceneContraRec                                                                 & 0.4979                                               & 0.4027                                                           & \underline{0.5448}                                                 & 0.4182                                                           & \underline{0.0762}                                                & \underline{0.0544}                                                & \underline{0.1045}                                                & \underline{0.0635}                                                & 0.1455  & \underline{0.1028}  & \underline{0.1983} & \underline{0.1199} \\ 
CARCA                                                             & \underline{0.5126}                                                           &  \underline{0.4373}                                                          & 0.5430                                                          & \underline{0.4472}                                                           & {\small OOM}  & {\small OOM}  & {\small OOM}  & {\small OOM} & {\small OOM}       &  {\small OOM}      & {\small OOM}        & {\small OOM}        \\
\hline
DSPnet--                                                                 & 0.5324	& 0.4612   &  0.5604	& 0.4703                                                            & 0.0742                                               & 0.0527                                                & 0.1047                                                & 0.0625                                                & 0.1443  & 0.1028  & 0.1987 & 0.1201 \\
DSPnet({\small w/o $\mathcal{L}_{\text{APR}}$, $\mathcal{L}_{\text{CCR}}$}) & 0.6115                                                           & 0.5292                                                           & 0.6625                                                           & 0.5459                                                           & 0.0843                                                          & 0.0617                                                          & 0.1123                                                          & 0.0707                                                          & 0.1680   & 0.1241  & 0.2206 & 0.1411 \\
DSPnet({\small w/o $\mathcal{L}_{\text{CCR}}$})                                & 0.6109                                                           & 0.5327                                                           & 0.6674                                                           & 0.5511                                                           & 0.0845                                                          & 0.0616                                                          & 0.1121                                                          & 0.0704                                                          & 0.1710  & 0.1266  & 0.2239 & \textbf{0.1437} \\
DSPnet({\small w/o $\mathcal{L}_{\text{APR}}$})                                & 0.6198                                                           & \textbf{0.5388}                                                           & 0.6684                                                           & \textbf{0.5545}                                                           & 0.0870                                                          & 0.0630                                                          & \textbf{0.1158}                                                          & 0.0723                                                          & 0.1711  & 0.1266  & 0.2229 & 0.1433 \\
\hline
\rowcolor{improvementblue} DSPnet                   & \begin{tabular}[c]{@{}c@{}}\textbf{0.6248}\\ {\small (+11.22\%)}\end{tabular} & \begin{tabular}[c]{@{}c@{}}0.5368\\ {\small (+9.95\%)}\end{tabular} & \begin{tabular}[c]{@{}c@{}}\textbf{0.6717}\\ {\small (+12.69\%)}\end{tabular} & \begin{tabular}[c]{@{}c@{}}0.5520\\ {\small (+10.48\%)}\end{tabular} &\begin{tabular}[c]{@{}c@{}}\textbf{0.0870}\\ {\small (+1.07\%)}\end{tabular} & \begin{tabular}[c]{@{}c@{}}\textbf{0.0632}\\ {\small (+0.88\%)}\end{tabular} & \begin{tabular}[c]{@{}c@{}}0.1155\\ {\small (+1.10\%)}\end{tabular} & \begin{tabular}[c]{@{}c@{}}\textbf{0.0725}\\ {\small (+0.90\%)}\end{tabular}         &\begin{tabular}[c]{@{}c@{}}\textbf{0.1712}\\ {\small (+2.53\%)}\end{tabular} & \begin{tabular}[c]{@{}c@{}}\textbf{0.1267}\\ {\small (+2.39\%)}\end{tabular} & \begin{tabular}[c]{@{}c@{}}\textbf{0.2240}\\ {\small (+2.57\%)}\end{tabular} & \begin{tabular}[c]{@{}c@{}}0.1436\\ {\small (+3.17\%)}\end{tabular}       \\ \hline
\end{tabular}
}}
\vspace{-6pt}
\end{table*}
\subsection{Overall Comparison}
In this section, we present the performance comparison for next item prediction task in Table~\ref{table:next_item_prediction}. \textbf{Given that next behavior can be stochastic while behavior distribution over a time period tends to be more stable, we introduce the period item prediction task, which focuses on forecasting user behaviors within a time period.} Results of the period item prediction task in Appendix~\ref{sec:appendix_more_experiments} Table~\ref{table:period_item_prediction}. 

From these tables, we observe that: 1) \textbf{Combining the scene information can obviously promote the modeling ability of sequential behaviors.} By incorporating this information, SceneContraRec improves its Recall@5 score from 0.3619 to 0.4979 on Outbrain. 2) \textbf{Compared to the public baselines, DSPnet is more capable of capturing the interplay between scenes and items in future behavior prediction.} Specifically, when examining the technique of integrating scene information, DSPnet shows a distinct advantage over other well-known methods (\textit{e.g.} SceneContraRec and CARCA), which rely solely on scene information as attribute embeddings. For example, DSPnet surpasses SceneContraRec by achieving an 11.22\% increase in Recall@5 on the Outbrain dataset. 
3) \textbf{The proposed dual sequence learning approach allows the model to be robust against user intention misalignment issue in scene-item behaviors, and learns better sequential dynamics.} We conclude this by comparing the results of DSPnet-- whose encoder is a one-to-one scene-item correspondence encoder, with DSPnet that incorporates a dual sequence learning encoder. 
4) \textbf{Comparing DSPnet and DSPnet(w/o $\mathcal{L}_{\text{CCR}}$), we see CCR loss enables the model to better learn representation invariance of historical sequences.} Note that CARCA contains a complex cross attention module and Bert4Rec involves the Cloze task that outputs large memory tensors for loss calculation. They have OOM issue on our large-scale datasets.

When removing $\mathcal{L}_{\text{APR}}$ and $\mathcal{L}_{\text{CCR}}$, we only have the vanilla dual sequence learning $\mathcal{L}_{\text{DSL}}^{\mathcal{V}}$ and $\mathcal{L}_{\text{DSL}}^{\mathcal{S}}$ in working. In this case, DSPnet(w/o $\mathcal{L}_{\text{APR}}$, $\mathcal{L}_{\text{CCR}}$) can still achieve better performance than public baselines, emphasizing the effectiveness of our dual sequence learning approach. Additionally, either removing $\mathcal{L}_{\text{APR}}$ or $\mathcal{L}_{\text{CCR}}$ would cause deteriorated results, because both objectives contribute to improved behavior prediction. 

\begin{figure}[t]
\centering
\begin{minipage}[t]{0.23\textwidth}
\centering
\includegraphics[width=\textwidth]{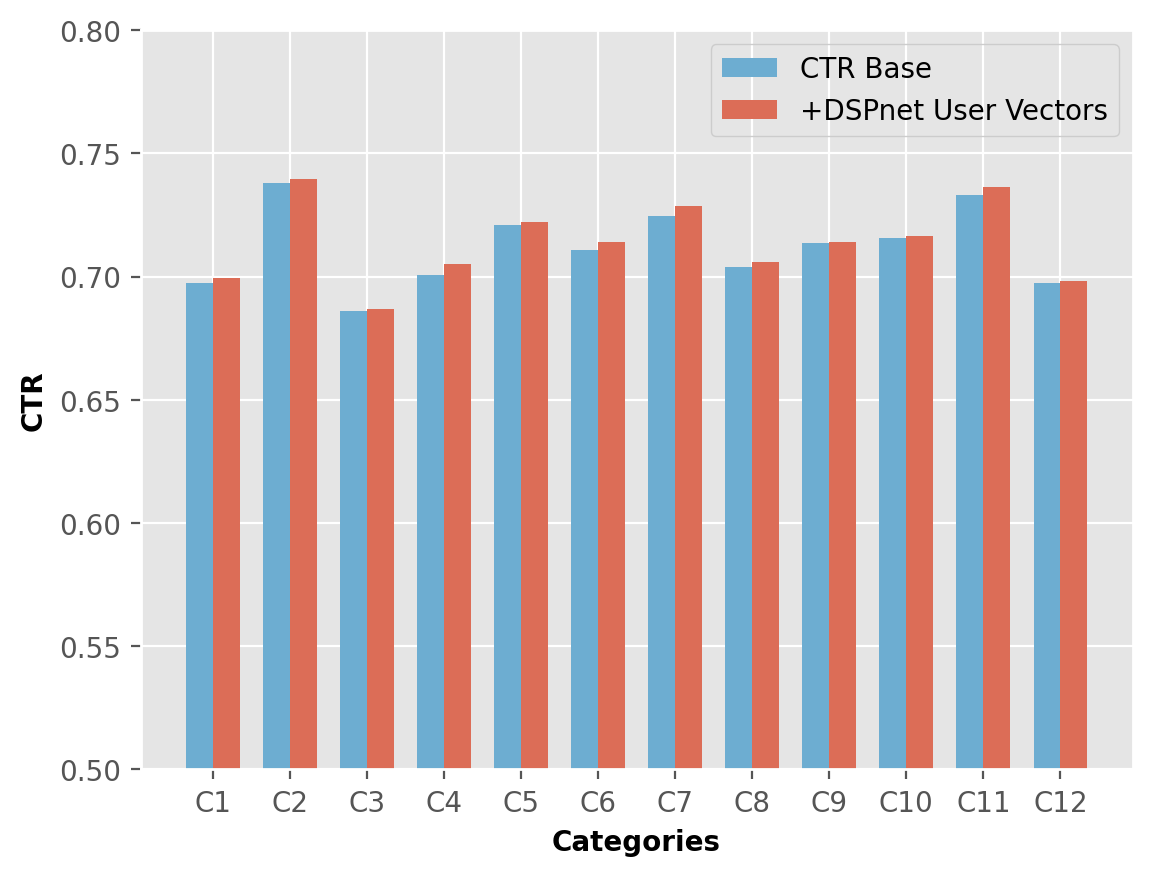}
\vspace{-20pt}
\caption*{{\small(a) CTR under Query Category}}
\end{minipage}
\begin{minipage}[t]{0.23\textwidth}
\centering
\includegraphics[width=\textwidth]{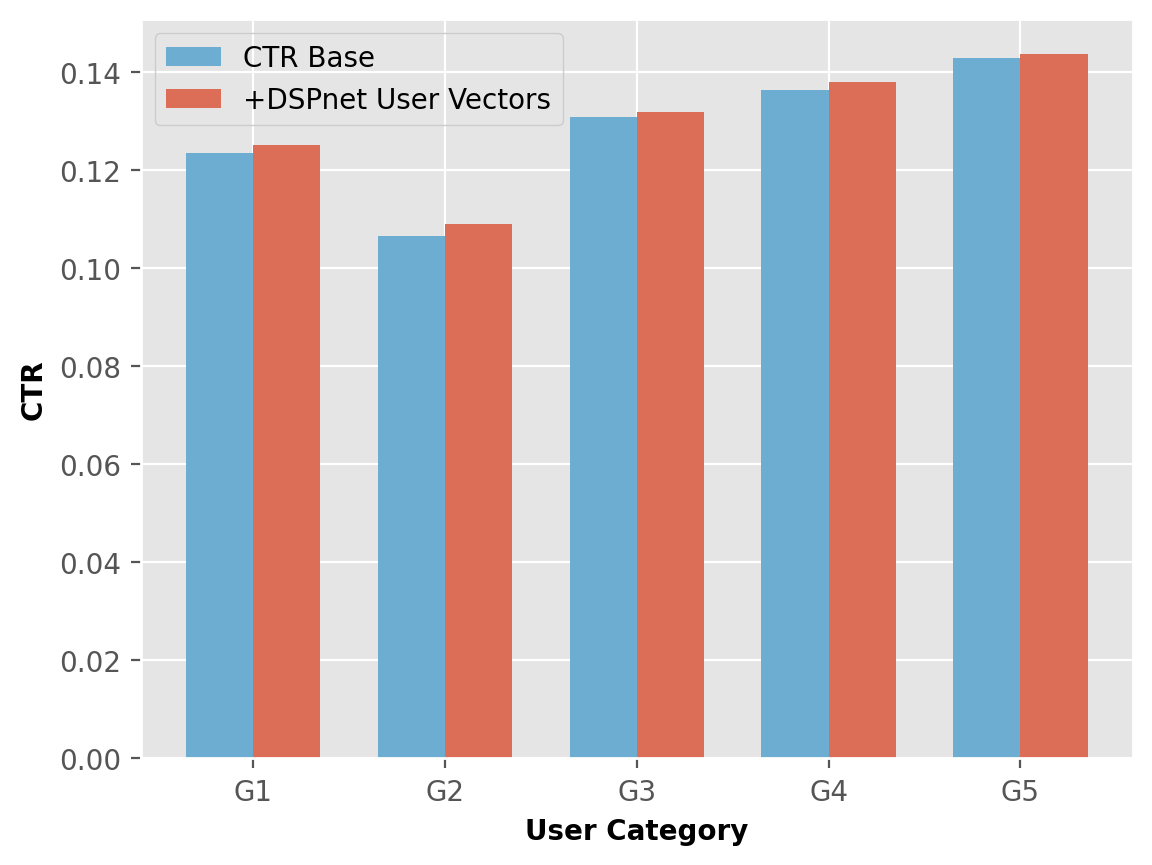}
\vspace{-20pt}
\caption*{\small{(b) CTR under User Category}}
\end{minipage}
\vspace{-12pt}
\caption{Online CTR comparison under different query categories and user categories of our CTR ranking model.}
\label{figure:online_AB}
\vspace{-10pt}
\end{figure}
\subsection{Online A/B Test and Deployment}
We further conduct an online A/B test in the product search system of our e-shopping app. In this experiment, we take DSPnet as a pretrained model on sequential user purchase behaviors and extract the learned user representations as additional user features for downstream click-through rate (CTR) prediction model. The baseline is our previously online serving model without this extracted user representations. The deployed DSPnet is trained on the collected AllScenePay dataset demonstrated in Section~\ref{sec:experiment_setup} by 32 A100 GPUs. In online inference, DSPnet utilizes 16 A100 GPUs and \textbf{takes the updated user sequence as input and outputs user representations daily.} 
The online evaluation metrics are real CTR, deal number and GMV (Gross Merchandise Value). 

When adding the user representations as additional features to the CTR model, we observed an absolute 0.04 point CTR increase, a relative 0.78\% deal growth and a relative 0.64\% GMV increase within 7-day online A/B test ranging from 2025-04-16 to 2025-04-21. In Figure~\ref{figure:online_AB}, we also present the CTR improvements across query and user categories. The figure clearly shows that adding DSPnet's user representations as input features can benefit downstream ranking performance. 
\textbf{Since May 2025, DSPnet has been fully deployed online to provide rich user features for our CTR ranking models, serving hundreds of millions of customers.}

\begin{table*}[]
\small
\centering
\caption{Study of the sequence feature enhancement module.}
\vspace{-6pt}
\label{table:feature_enhancement}
\renewcommand{\arraystretch}{0.95}
 \setlength{\tabcolsep}{0.5mm}{ 
  \scalebox{0.9}{
\begin{tabular}{c|cccc|cccc|cccc}
\hline
Dataset               & \multicolumn{4}{c|}{Ourbrain}     & \multicolumn{4}{c|}{AllScenePay-1m} & \multicolumn{4}{c}{AllScenePay-10m} \\ \hline
Model                 & R@5    & N@5    & R@10   & N@10   & R@5     & N@5     & R@10   & N@10   & R@5    & N@5    & R@10    & N@10    \\ \hline
w/o concat            & 0.3661 & 0.2997 & 0.4028 & 0.3118 & 0.0701  & 0.0488  & 0.0994 & 0.0582 & 0.1412	&0.1000	&0.1953	&0.1175    \\
w/o MLP               & 0.5400          & 0.4719          & 0.5704          & 0.4819          & \textbf{0.0899} & 0.0627          & \textbf{0.1215} & \textbf{0.0729} & 0.1645  & 0.1201  & 0.2184 & 0.1375 \\ \hline
DSPnet(MLP\_layers=1) & 0.5633          & 0.4795          & 0.6211          & 0.4984          & 0.0868          & 0.0634          & 0.1138          & 0.0720          & 0.1676  & 0.1241  & 0.2194 & 0.1409 \\
DSPnet(MLP\_layers=3) & 0.6175          & 0.5360          & 0.6644          & 0.5513          & 0.0857          & 0.0623          & 0.1165          & 0.0723          & 0.1621  & 0.1207  & 0.2111 & 0.1365 \\ \hline
DSPnet(MLP\_layers=2) & \textbf{0.6248} & \textbf{0.5368} & \textbf{0.6717} & \textbf{0.5520}  & 0.0870  & \textbf{0.0632}  & 0.1155 & 0.0725 & \textbf{0.1712}  & \textbf{0.1267}  & \textbf{0.2240} & \textbf{0.1436}        \\ \hline
\end{tabular}
}}
\vspace{-6pt}
\end{table*}
\subsection{Study of Sequence Feature Enhancement}
In our dual sequence learning, sequence feature enhancement is an important module to capture the interplay between two historical sequences. We conduct an experiment to study how this component influences model performance. 
Results are given in Table~\ref{table:feature_enhancement}. 

From this table, we can conclude that: 1) \textbf{The removal of feature enhancement module (denoted as ``w/o concat") leads to an obvious decrease in model performance}, underscoring the crucial role of sequence feature enhancement module in capturing the interplay and dynamics for predicting future behaviors.
2) When the MLP layers are excluded (as indicated by ``w/o MLP"), the model relies solely on concatenation operation to integrate information. This limitation results in poorer performance compared to the variants that include MLP layers, as the ``w/o MLP" variant lacks capacity and flexibility to fuse user representations.
3) Different number of MLP layers lead to different model performances. With proper MLP layers, we can enhance the model's capability, allowing for better interplay modeling.

\begin{figure*}[t]
\centering
\begin{minipage}[t]{0.23\textwidth}
\centering
\includegraphics[width=\textwidth]{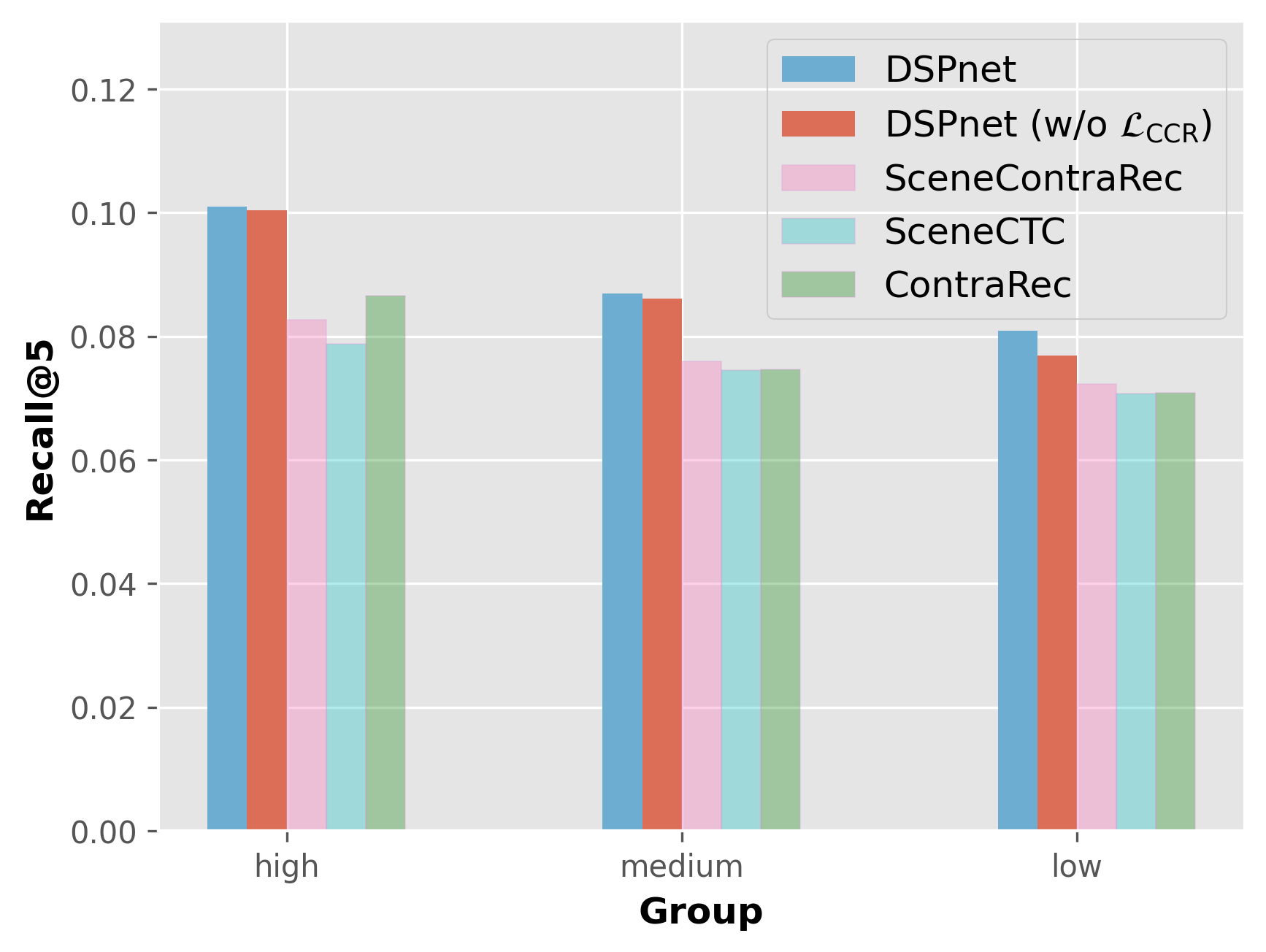}
\vspace{-20pt}
\caption*{(a)}
\end{minipage}
\begin{minipage}[t]{0.23\textwidth}
\centering
\includegraphics[width=\textwidth]{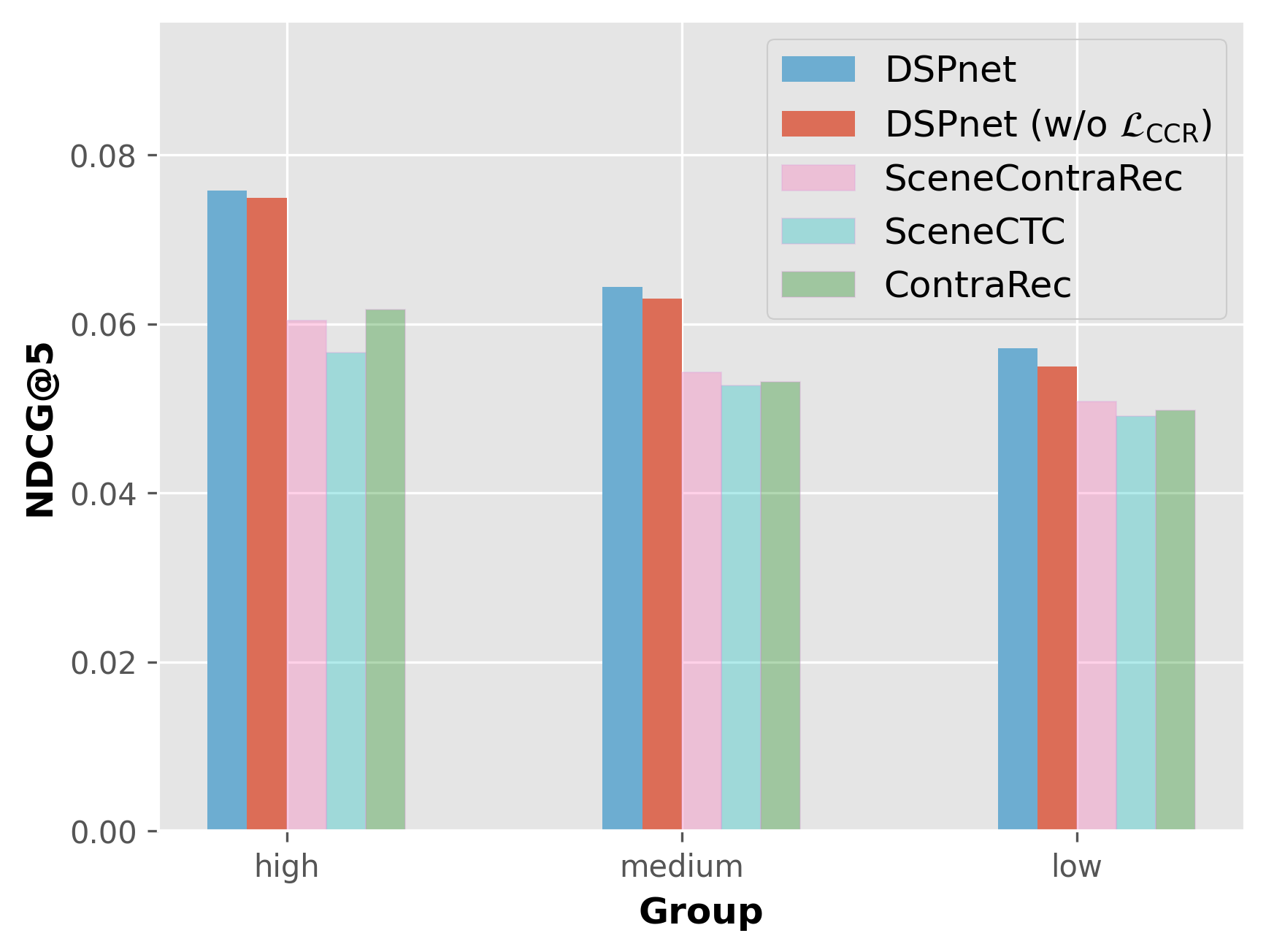}
\vspace{-20pt}
\caption*{(b)}
\end{minipage}
\begin{minipage}[t]{0.23\textwidth}
\centering
\includegraphics[width=\textwidth]{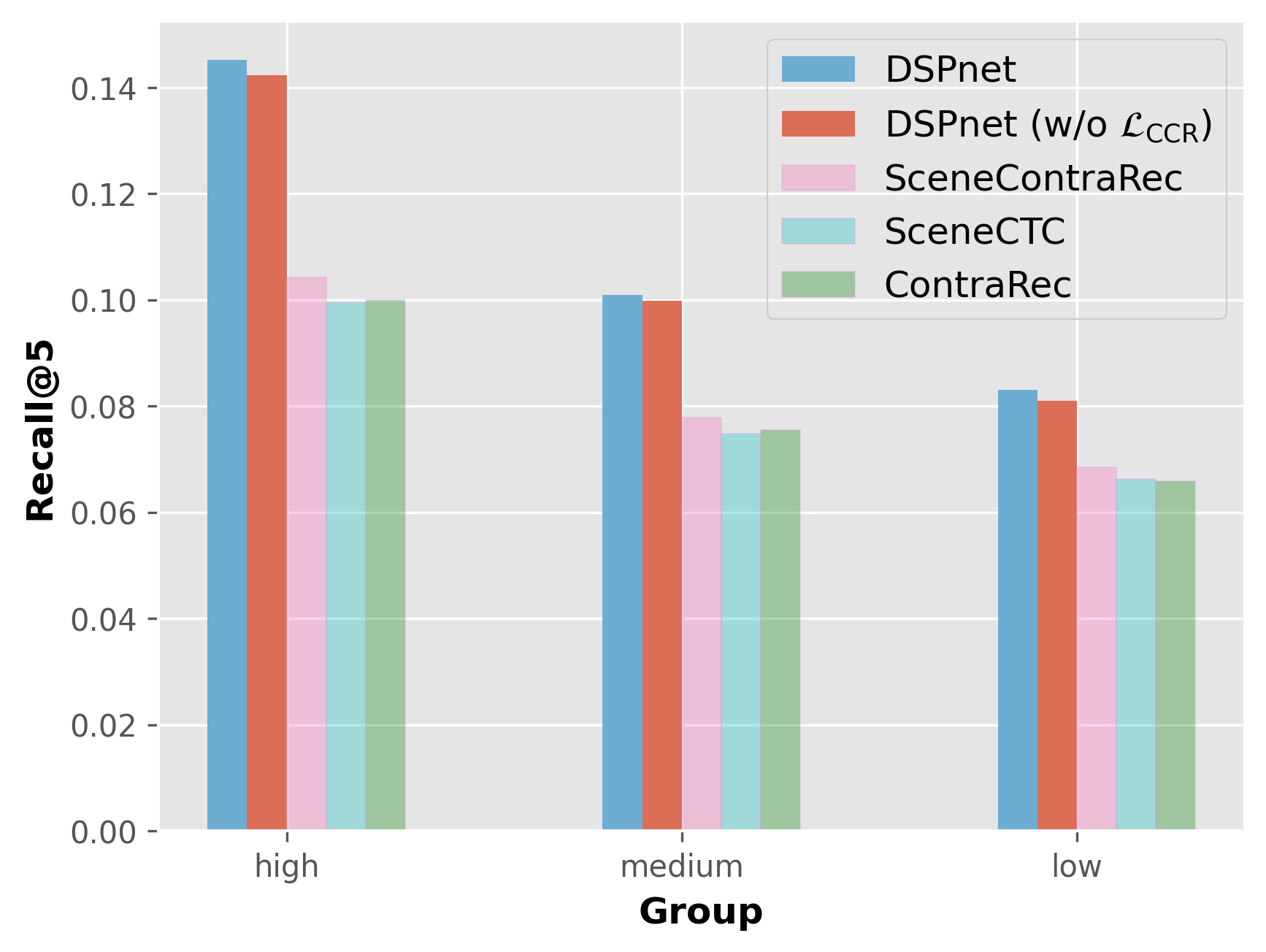}
\vspace{-20pt}
\caption*{(c)}
\end{minipage}
\begin{minipage}[t]{0.23\textwidth}
\centering
\includegraphics[width=\textwidth]{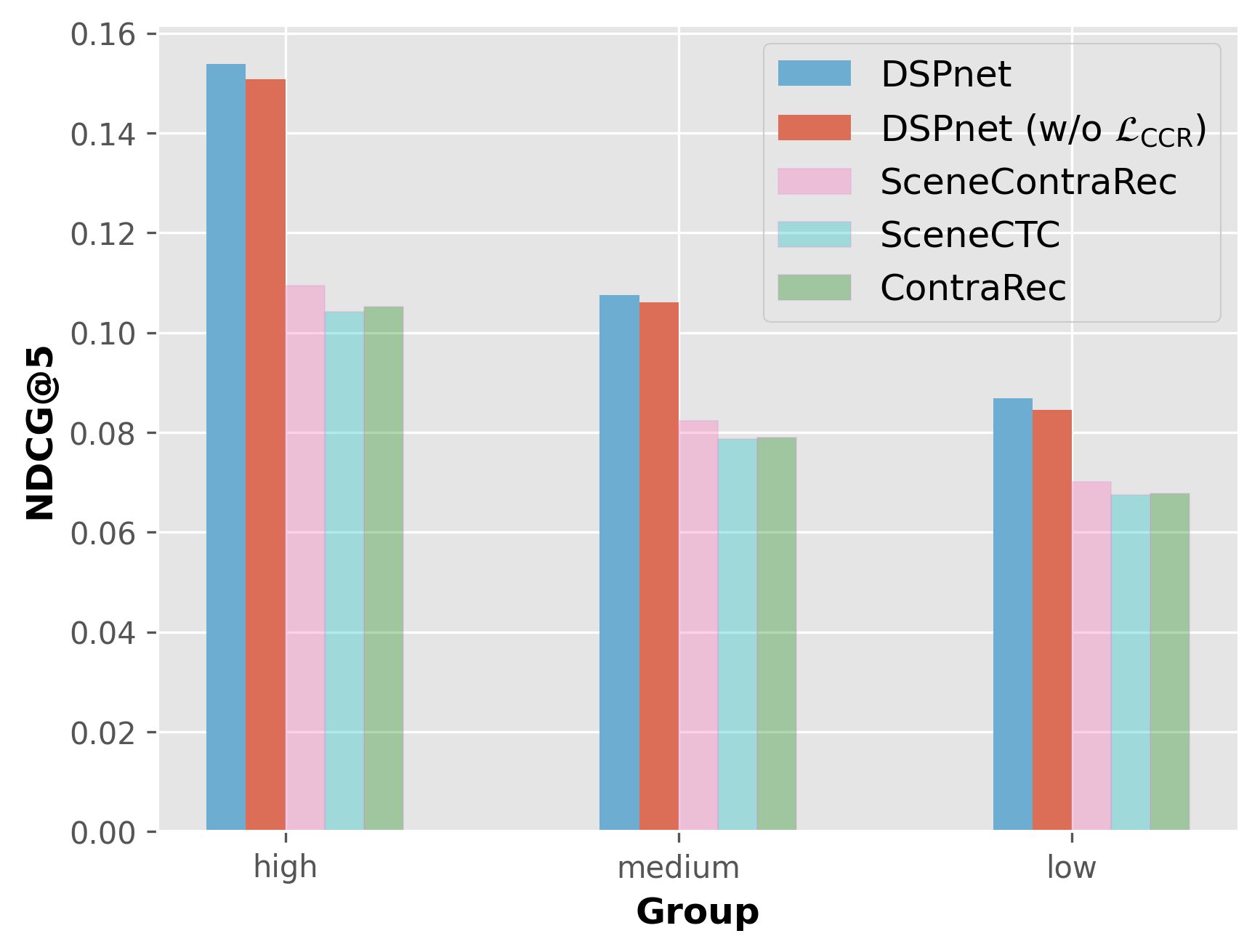}
\vspace{-20pt}
\caption*{(d)}
\end{minipage}
\vspace{-8pt}
\caption{Performance comparison of different methods on different user categories. (a) and (b) indicate results on next item prediction task. (c) and (d) show results on period item prediction task.}
\label{figure:user_group}
\vspace{-8pt}
\end{figure*}
\subsection{Performance on Different User Categories}
As the length of user sequences usually follows a severe long-tailed distribution, we conduct an experiment to study the model's generalization ability on different parts of the distribution. We split test user sequences into three groups (\textit{i.e.} ``high", ``medium", ``low") based on sequence lengths. Results are given in Figure~\ref{figure:user_group}.

From this figure, we see \textbf{DSPnet has consistent improvements on different user categories over baselines}. This demonstrates the superior generalization capability of our method.
Moreover, DSPnet shows a more significant performance gap compared to DSPnet (w/o $\mathcal{L}_{\text{CCR}}$) in the ``low'' group than in the ``medium'' group.
As discussed in Section~\ref{sec:CCR}, CCR loss has the advantage of considering relationships inner positives or negatives, which is important for skewed data distributions. \textbf{To highlight the advantages of CCR loss over conventional contrastive loss, we replaced the contrastive regularization in SceneContraRec with CCR loss. See Appendix~\ref{append:appen_combineCCR} for results and analysis.}


\begin{figure*}[t]
\centering
\begin{minipage}[t]{0.23\textwidth}
\centering
\includegraphics[width=\textwidth]{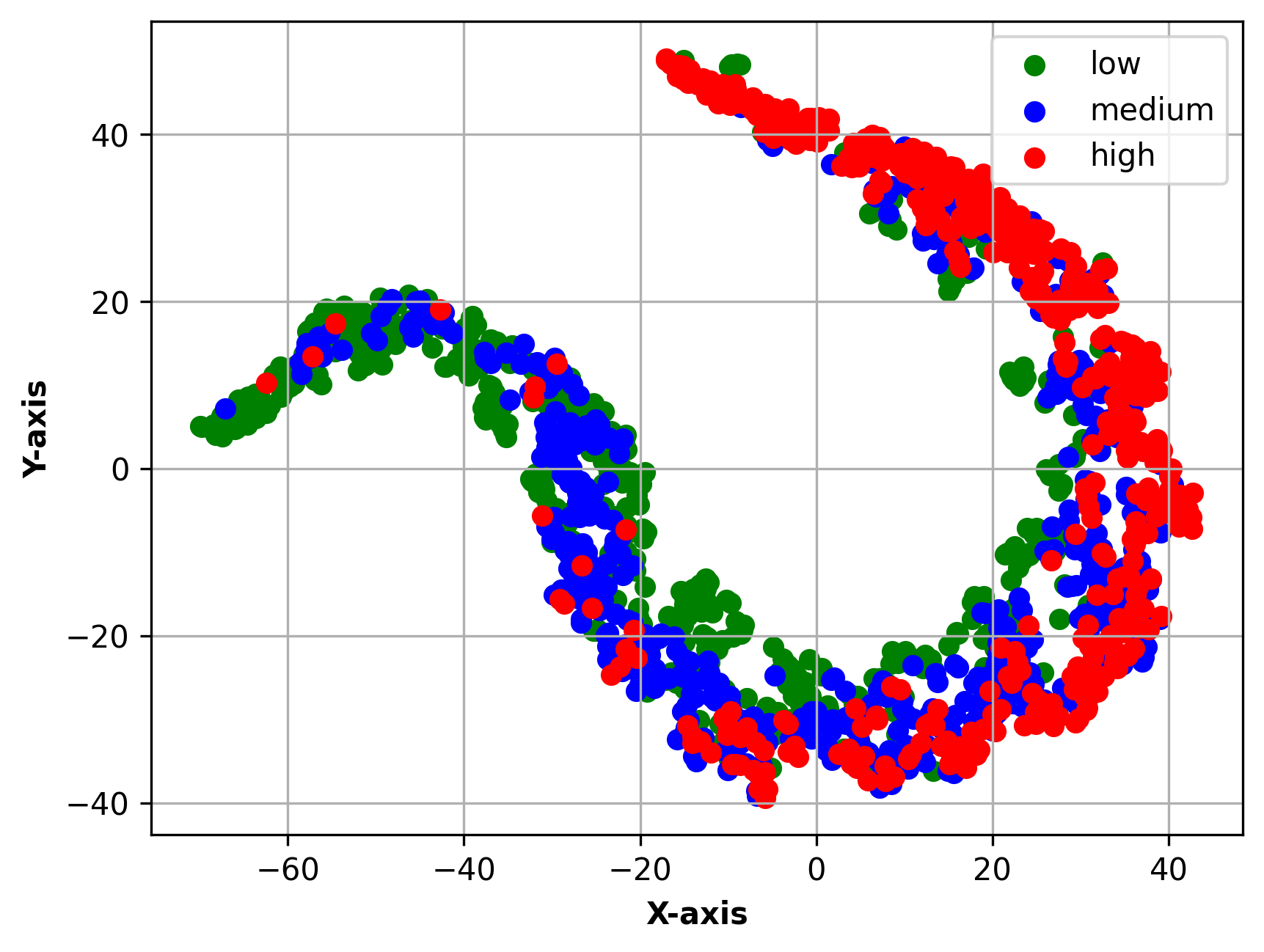}
\vspace{-20pt}
\caption*{{\small(a) ContraRec}}
\end{minipage}
\begin{minipage}[t]{0.23\textwidth}
\centering
\includegraphics[width=\textwidth]{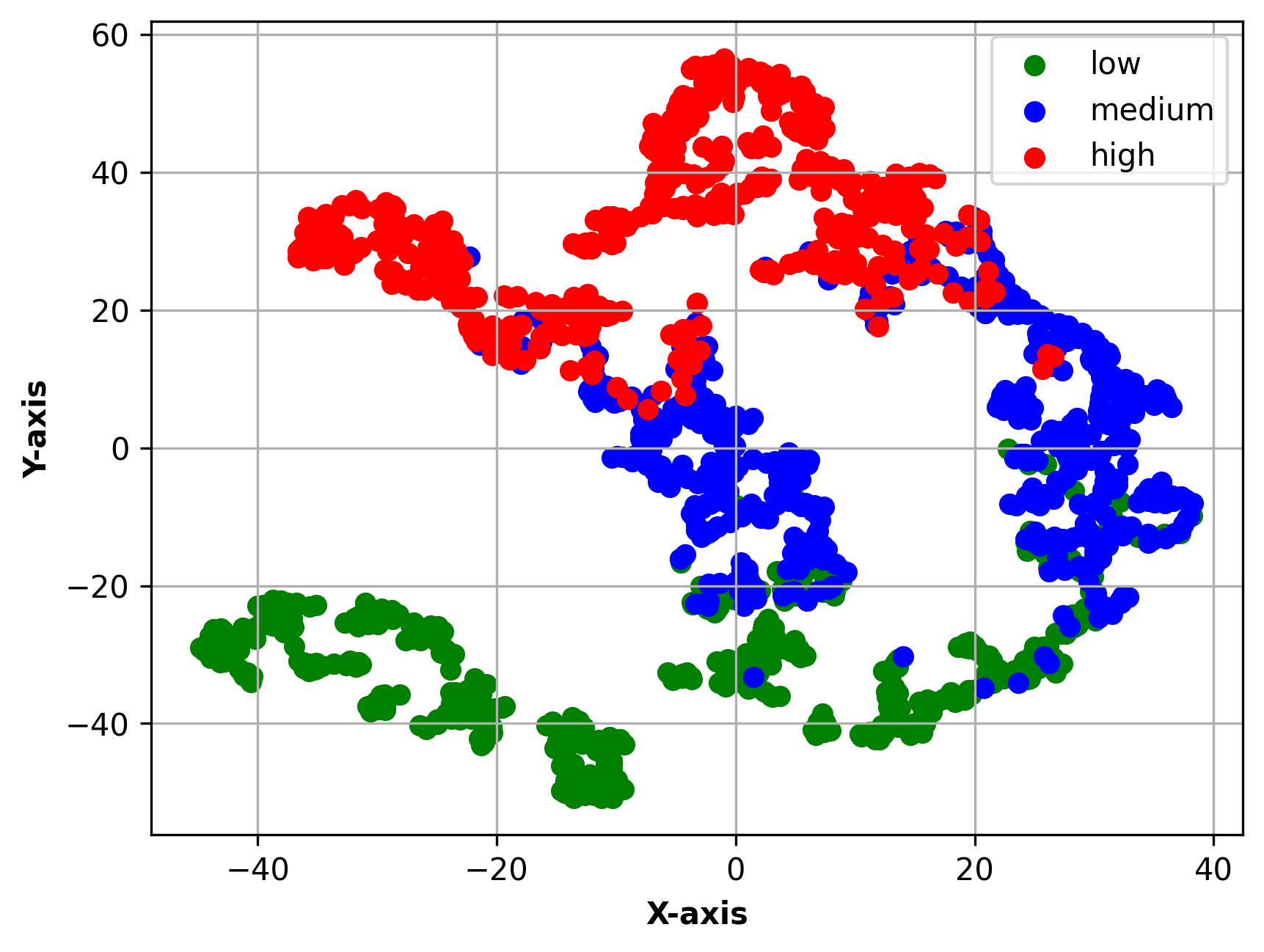}
\vspace{-20pt}
\caption*{\small{(b) SceneContraRec}}
\end{minipage}
\begin{minipage}[t]{0.23\textwidth}
\centering
\includegraphics[width=\textwidth]{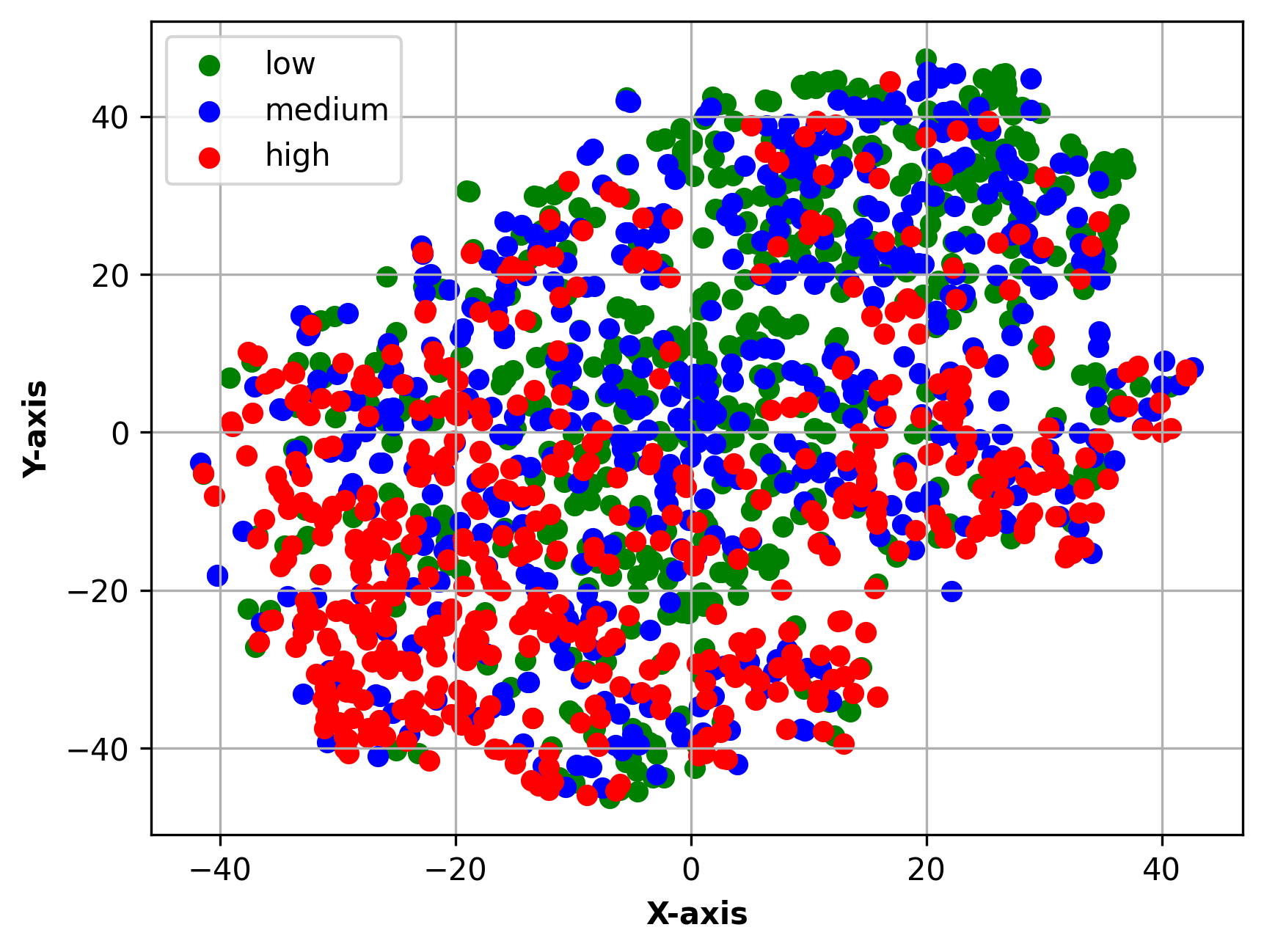}
\vspace{-20pt}
\caption*{\small{(c) DSPnet(w/o $\mathcal{L}_{\text{CCR}}$)}}
\end{minipage}
\begin{minipage}[t]{0.23\textwidth}
\centering
\includegraphics[width=\textwidth]{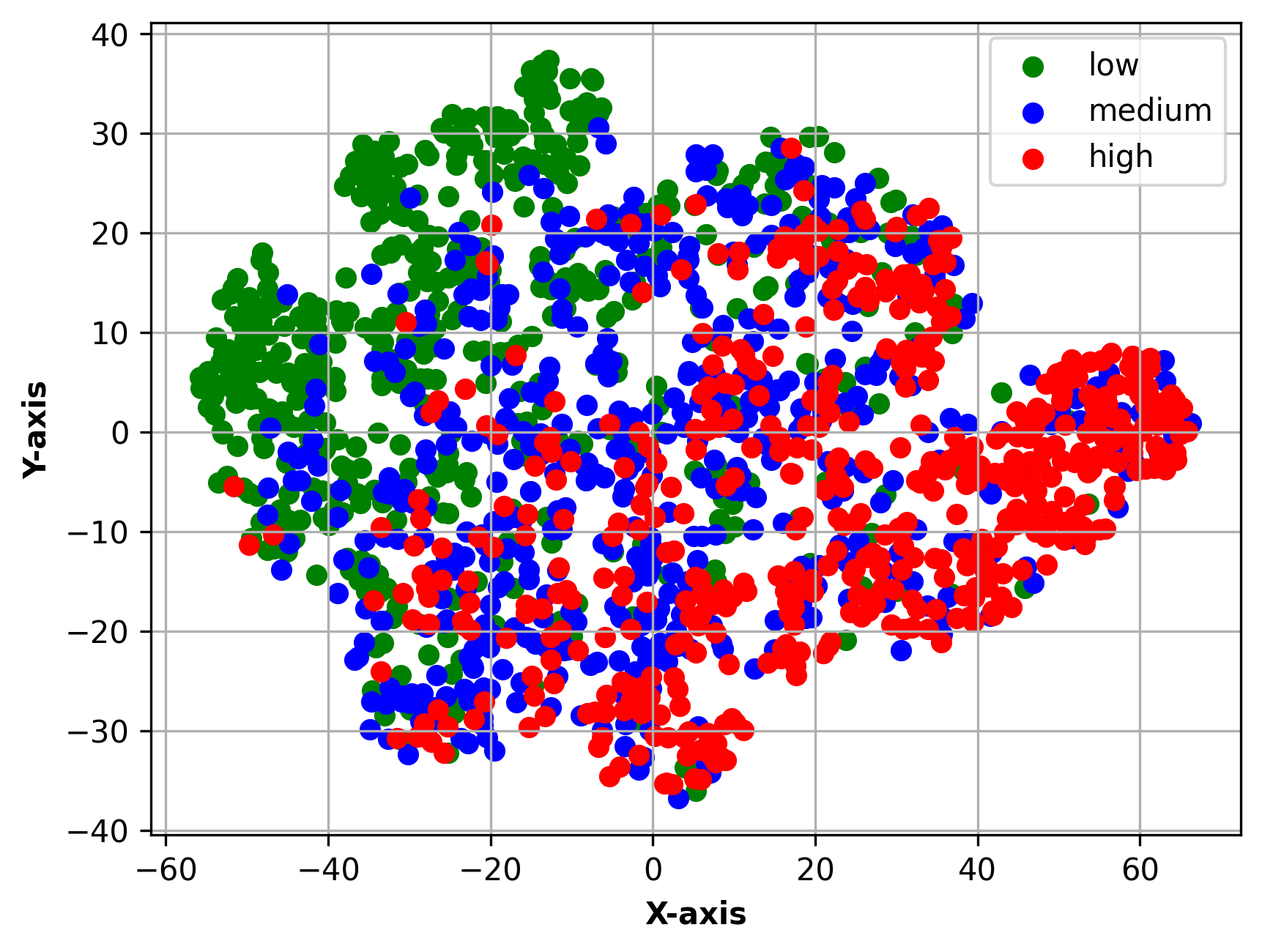}
\vspace{-20pt}
\caption*{\small{(d) DSPnet}}
\end{minipage}
\vspace{-8pt}
\caption{The t-SNE visualization of learned user representations on AllScenePay-1m dataset.}
\label{figure:user_tsne}
\vspace{-10pt}
\end{figure*}
\subsection{Visualization of User Representations}
In sequential behavior modeling, user representations are usually encoded from historical behaviors and largely influence the performance of final behavior prediction. We here investigate whether the learned user representations are better than baselines.
Specifically, we obtain user representations of test set and split them into three groups (\textit{i.e.} ``high", ``medium", ``low") based on their number of historical interactions. Then, for each group, we randomly sample 500 user representations for t-SNE visualization. The results of different methods are given in Figure~\ref{figure:user_tsne}.

From this figure, we summarize that: 1) User representations generated by DSPnet achieve significant improvements compared to baselines. In ContraRec, representations from different groups are intertwined, leading to less discrimination. SceneContraRec manages to classify representations well, but they tend to converge into a small region, which can be detrimental for personalization in subsequent recommendation or retrieval tasks. Additionally, the representation distances in SceneContraRec do not align with the expectation that distance between ``high" and ``low" groups should be greater than that between ``medium" and ``low" groups. In contrast, \textbf{DSPnet's representations are easily distinguishable, do not collapse into a small subspace, and exhibit clear distance interpretability.} 2) When comparing panels (c) and (d), it is evident that the inclusion of CCR loss facilitates the learning of more compact representations, particularly within the ``low" group. CCR loss promotes representation invariance and generalization abilities on skewed user behaviors. \textbf{More experiments and analysis are provided in the Appendix.}

\section{Conclusion and Future Work}
Leveraging the defined scene feature for sequential user behavior modeling is a crucial but less studied issue in recent works. 
In this study, we propose a novel framework called DSPnet that effectively captures the interplay between historical scene and item sequences, while being robust against user intention misalignment issue of scene-item data. Additionally, recognizing the randomness and noise inherent in user behaviors, we introduce CCR loss to enhance representation invariance, thereby improving the learning of dynamic interests. Through both theoretical analysis and empirical evaluation, we demonstrate that DSPnet does better at modeling sequential user behaviors and exhibits superior performance in skewed data scenarios.

Although DSPnet has achieved remarkable performance of user behavior prediction, it still has certain limitations. 
For instance, it only incorporates scene and item sequences, leaving potential to improve performance by integrating additional feature information (\textit{e.g.} item category, behavior type). We will incorporate more rich features into DSPnet in future work.


\newpage
\bibliographystyle{ACM-Reference-Format}
\bibliography{sample-base}

\appendix

\section{Derivation of the Joint Log-Likelihood}
\label{appendix:joint_log_likelihood}
Let $\boldsymbol{v}$ and $\boldsymbol{s}$ denote the observed interacted item and scene of user $\boldsymbol{u}$, respectively. Then, the joint log-likelihood is composed of a sum over the likelihoods of individual data points $\sum_{\boldsymbol{u}} [\log p_{\theta}(\boldsymbol{v},\boldsymbol{s})]$, where $p_{\theta}(\boldsymbol{v},\boldsymbol{s})$ is the probability density function.
Given the observed item-scene behaviors $(\boldsymbol{v},\boldsymbol{s})$, we denote $\mathcal{V}$ and $\mathcal{S}$ as the historically interacted items and scenes sequentially before $\boldsymbol{v}$ and $\boldsymbol{s}$, respectively. 
The corresponding encoded latent representations of $\mathcal{V}$ and $\mathcal{S}$ are denoted as $\boldsymbol{z}_\mathcal{V}$ and $\boldsymbol{z}_\mathcal{S}$, respectively.
Then, drawing from the idea of maximizing the marginal log-likelihood in Variational Autoencoders (VAEs)~\cite{kingma2013auto}, $\log p_{\theta}(\boldsymbol{v},\boldsymbol{s})$ is written as:
\begin{small}
\begin{align}
    \log p_{\theta}(\boldsymbol{v},\boldsymbol{s})=D_{KL}[q_{\phi}(\boldsymbol{z}_{\mathcal{V}},\boldsymbol{z}_{\mathcal{S}}|\mathcal{V}, \mathcal{S})||p(\boldsymbol{z}_{\mathcal{V}},\boldsymbol{z}_{\mathcal{S}}|\mathcal{V}, \mathcal{S})] +\mathcal{L}_{\text{ELBO}},
\end{align} 
\end{small}

where the first term denotes KL divergence between parameterized posterior $q_{\phi}(\boldsymbol{z}_{\mathcal{V}},\boldsymbol{z}_{\mathcal{S}}|\mathcal{V}, \mathcal{S})$ and the true one $p(\boldsymbol{z}_{\mathcal{V}},\boldsymbol{z}_{\mathcal{S}}|\mathcal{V}, \mathcal{S})$. This KL divergence is non-negative, so the second term is the \textit{evidence lower bound (ELBO)} on the log-likelihood $\log p_{\theta}(\boldsymbol{v},\boldsymbol{s})$. 

Following the derivation in VAE~\cite{kingma2013auto}, when maximizing the above joint log-likelihood, we can maximize the following \textit{ELBO}:
\begin{align}
\label{eq:elbo_appendix}
    \max_{\theta,\phi} \mathcal{L}_{\text{ELBO}}=&\underbrace{\mathbb{E}_{q_{\phi}(\boldsymbol{z}_{\mathcal{V}},\boldsymbol{z}_{\mathcal{S}}|\mathcal{V},\mathcal{S})}[\log p_{\theta}(\boldsymbol{v},\boldsymbol{s}|\boldsymbol{z}_{\mathcal{V}},\boldsymbol{z}_{\mathcal{S}})]}_{\text{encoder-decoder}} \nonumber \\
    &\underbrace{- D_{KL}[q_{\phi}(\boldsymbol{z}_{\mathcal{V}},\boldsymbol{z}_{\mathcal{S}}|\mathcal{V},\mathcal{S})||p(\boldsymbol{z}_{\mathcal{V}},\boldsymbol{z}_{\mathcal{S}})]}_{\text{joint prior regularization}},
\end{align}
where $p_{\theta}(\boldsymbol{v},\boldsymbol{s}|\boldsymbol{z}_{\mathcal{V}},\boldsymbol{z}_{\mathcal{S}})$ is the conditional distribution parameterized by $\theta$. The first term actually shows an encoder-decoder architecture, while the second term indicates a joint prior regularization on $q_{\phi}(\boldsymbol{z}_{\mathcal{V}},\boldsymbol{z}_{\mathcal{S}}|\mathcal{V},\mathcal{S})$.


\textbf{The Encoder-Decoder}: Given $\mathcal{V}$ and $\mathcal{S}$, the encoded latent representations $\boldsymbol{z}_{\mathcal{V}}$ and $\boldsymbol{z}_{\mathcal{S}}$ are conditional independent, and the posterior can be written as:
\begin{align}
    q_{\phi}(\boldsymbol{z}_{\mathcal{V}},\boldsymbol{z}_{\mathcal{S}}| \mathcal{V},\mathcal{S})&=q_{\phi_1}(\boldsymbol{z}_{\mathcal{V}}|\boldsymbol{z}_{\mathcal{S}},\mathcal{V},\mathcal{S})q_{\phi_2}(\boldsymbol{z}_{\mathcal{S}}|\mathcal{V},\mathcal{S}) \nonumber \\
    \label{eq:posterior_2}
    &=q_{\phi_1}(\boldsymbol{z}_{\mathcal{V}}|\mathcal{V},\mathcal{S})q_{\phi_2}(\boldsymbol{z}_{\mathcal{S}}|\mathcal{V},\mathcal{S}),
\end{align}
which indicates both the representations of historical items and scenes are not solely dependent from their own sequences. Instead, these representations are derived from $\mathcal{V}$ and $\mathcal{S}$, indicating that the representation of historical items $\boldsymbol{z}_{\mathcal{V}}$ is influenced by the contextual scene sequence, similarly influencing $\boldsymbol{z}_{\mathcal{S}}$.

Similarly, given $\boldsymbol{z}_{\mathcal{V}}$ and $\boldsymbol{z}_{\mathcal{S}}$, $\boldsymbol{v}$ and $\boldsymbol{s}$ are conditional independent, the conditional distribution is written as:
\begin{align}
    \label{eq:conditoal_ori}
    p_{\theta}(\boldsymbol{v},\boldsymbol{s}|\boldsymbol{z}_{\mathcal{V}},\boldsymbol{z}_{\mathcal{S}})&=p_{\theta_1}(\boldsymbol{v}|\boldsymbol{s},\boldsymbol{z}_{\mathcal{V}},\boldsymbol{z}_{\mathcal{S}})p_{\theta_2}(\boldsymbol{s}|\boldsymbol{z}_{\mathcal{V}},\boldsymbol{z}_{\mathcal{S}}) \nonumber \\
    &=p_{\theta_1}(\boldsymbol{v}|\boldsymbol{z}_{\mathcal{V}},\boldsymbol{z}_{\mathcal{S}}) p_{\theta_2}(\boldsymbol{s}|\boldsymbol{z}_{\mathcal{V}},\boldsymbol{z}_{\mathcal{S}}),
\end{align}
which indicates we employ both the information from $\boldsymbol{z}_{\mathcal{V}}$ and $\boldsymbol{z}_{\mathcal{S}}$ to make the individual prediction of $\boldsymbol{v}$ and $\boldsymbol{s}$.

\textbf{The Joint Prior Regularization}: 
The second term in Eq.~\ref{eq:elbo_appendix} represents a joint prior on the posterior $q_{\phi}(\boldsymbol{z}_{\mathcal{V}},\boldsymbol{z}_{\mathcal{S}}|\mathcal{V},\mathcal{S})$ for $\boldsymbol{z}_{\mathcal{V}}$ and $\boldsymbol{z}_{\mathcal{S}}$. Given the complexity of the joint prior $p(\boldsymbol{z}_{\mathcal{V}},\boldsymbol{z}_{\mathcal{S}})$, we simplify its implementation by assuming $p(\boldsymbol{z}_{\mathcal{V}},\boldsymbol{z}_{\mathcal{S}})=p(\boldsymbol{z}_{\mathcal{V}})p(\boldsymbol{z}_{\mathcal{S}})$. This choice aligns with recent works~\cite{xu_towards2023,tomczak2017vae}, allowing for a more straightforward and efficient implementation. By integrating this with Eq.~\ref{eq:posterior_2}, the joint prior regularization (Eq.~\ref{eq:elbo_appendix}) can be formulated as:
\begin{small}
\begin{align}
    &D_{KL}[q_{\phi}(\boldsymbol{z}_{\mathcal{V}},\boldsymbol{z}_{\mathcal{S}}|\mathcal{V},\mathcal{S})||p(\boldsymbol{z}_{\mathcal{V}},\boldsymbol{z}_{\mathcal{S}})] \nonumber \\ 
    &=D_{KL}[q_{\phi}(\boldsymbol{z}_{\mathcal{V}}|\mathcal{V},\mathcal{S})q_{\phi}(\boldsymbol{z}_{\mathcal{S}}|\mathcal{V},\mathcal{S})||p(\boldsymbol{z}_{\mathcal{V}})p(\boldsymbol{z}_{\mathcal{S}})]\label{eq:prior_2} \nonumber \\
    &=D_{KL}[q_{\phi}(\boldsymbol{z}_{\mathcal{V}}|\mathcal{V},\mathcal{S})||p(\boldsymbol{z}_{\mathcal{V}})]+D_{KL}[q_{\phi}(\boldsymbol{z}_{\mathcal{S}}|\mathcal{V},\mathcal{S})||p(\boldsymbol{z}_{\mathcal{S}})]. 
\end{align} 
\end{small}

It is worthwhile to point out that the joint prior assumption $p(\boldsymbol{z}_{\mathcal{V}},\boldsymbol{z}_{\mathcal{S}})=p(\boldsymbol{z}_{\mathcal{V}})p(\boldsymbol{z}_{\mathcal{S}})$ is not a perfect choice. In future work, we may explore more intricate joint priors, leveraging the insights from~\cite{tomczak2017vae,rezende2015variational,yin2018semi}.

\textbf{Rewrite the $\mathcal{L}_{\text{ELBO}}$}: By integrating Eq.~\ref{eq:posterior_2}, Eq.~\ref{eq:conditoal_ori} and Eq.~\ref{eq:prior_2} together, we can rewrite the \textit{ELBO} in Eq.~\ref{eq:elbo_appendix} as follows:
\begin{small}
\begin{align}
  \label{eq:new_elbo}
    &\max_{\theta_1,\theta_2,\phi_1,\phi_2} \mathcal{L}_{\text{ELBO}} = \nonumber \\
    &\mathbb{E}_{q_{\phi_1}(\boldsymbol{z}_{\mathcal{V}}|\mathcal{V},\mathcal{S})q_{\phi_2}(\boldsymbol{z}_{\mathcal{S}}|\mathcal{V},\mathcal{S})}[\log p_{\theta_1}(\boldsymbol{v}|\boldsymbol{z}_{\mathcal{V}},\boldsymbol{z}_{\mathcal{S}})p_{\theta_2}(\boldsymbol{s}|\boldsymbol{z}_{\mathcal{V}},\boldsymbol{z}_{\mathcal{S}})] \nonumber \\
    & - D_{KL}[q_{\phi_1}(\boldsymbol{z}_{\mathcal{V}}|\mathcal{V},\mathcal{S})||p(\boldsymbol{z}_{\mathcal{V}})] - D_{KL}[q_{\phi_2}(\boldsymbol{z}_{\mathcal{S}}|\mathcal{V},\mathcal{S})||p(\boldsymbol{z}_{\mathcal{S}})].  
\end{align}   
\end{small}

\textbf{The Objective Function}: Maximizing the above ELBO is equivalent to minimizing its negative version. We summarize the optimization objective function as:
\begin{align}
    \min_{\theta_1,\theta_2,\phi_1,\phi_2} \mathcal{L} =&-\mathbb{E}_{q_{\phi_1}(\boldsymbol{z}_{\mathcal{V}}|\mathcal{V},\mathcal{S})}[\log p_{\theta_1}(\boldsymbol{v}|\boldsymbol{z}_{\mathcal{V}},\boldsymbol{z}_{\mathcal{S}})] \nonumber\\
    &- \mathbb{E}_{q_{\phi_2}(\boldsymbol{z}_{\mathcal{S}}|\mathcal{V},\mathcal{S})}[\log p_{\theta_2}(\boldsymbol{s}|\boldsymbol{z}_{\mathcal{V}},\boldsymbol{z}_{\mathcal{S}})] \nonumber \\
    & +D_{KL}[q_{\phi_1}(\boldsymbol{z}_{\mathcal{V}}|\mathcal{V},\mathcal{S})||p(\boldsymbol{z}_{\mathcal{V}})] \nonumber \\
    &+ D_{KL}[q_{\phi_2}(\boldsymbol{z}_{\mathcal{S}}|\mathcal{V},\mathcal{S})||p(\boldsymbol{z}_{\mathcal{S}})].
\end{align}

where the first and second term indicate we obtain the latent representations from historical sequences $\mathcal{V},\mathcal{S}$, and then we use them to predict the future item $\boldsymbol{v}$ and scene $\boldsymbol{s}$. The third and fourth term show prior regularization on the latent representations, which can be implemented by adversarial learning shown in~\cite{makhzani2015adversarial}. \textbf{Therefore, we can see that the objective function actually is equivalent to our dual sequence framework without specifying the detailed encoder-decoder networks and prediction loss.}

\begin{figure}[t]
\centering
\begin{minipage}[t]{0.23\textwidth}
\centering
\includegraphics[width=\textwidth]{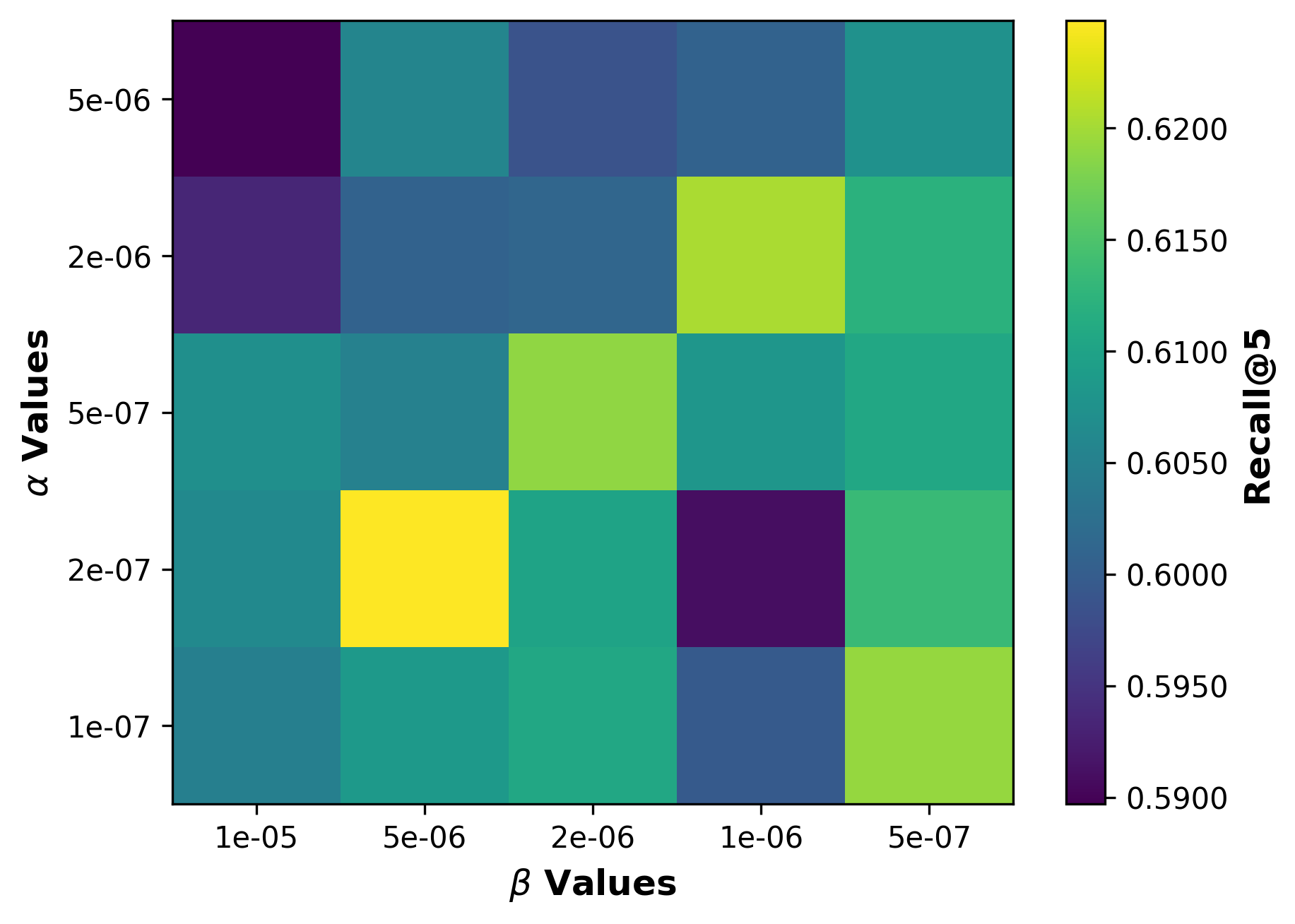}
\vspace{-20pt}
\caption*{(a){\small Outbrain}}
\end{minipage}
\begin{minipage}[t]{0.23\textwidth}
\centering
\includegraphics[width=\textwidth]{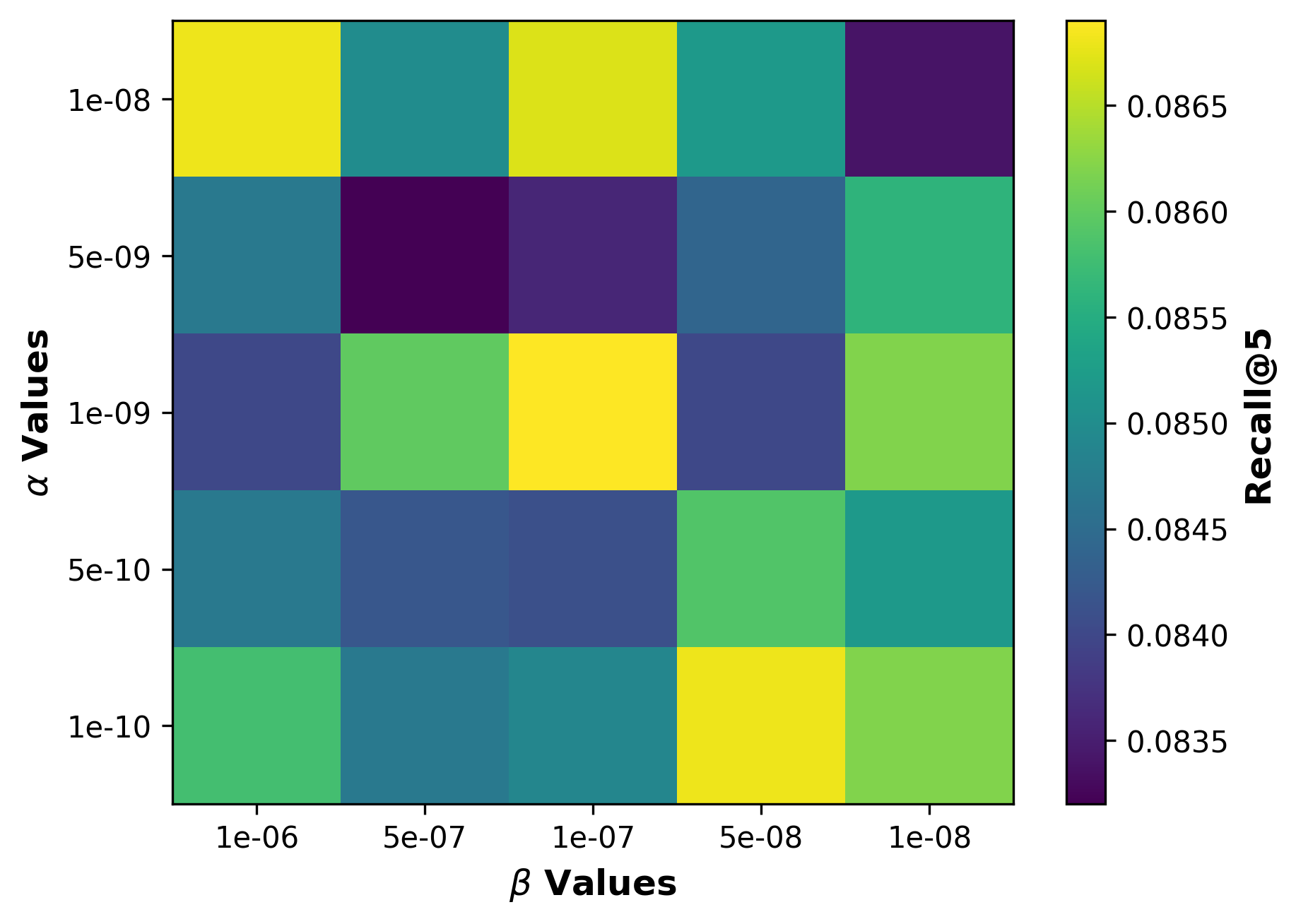}
\vspace{-20pt}
\caption*{(b){\small AllScenePay-1m}}
\end{minipage}
\vspace{-8pt}
\caption{The performance of different hyper-parameters.}
\label{figure:hyper_params}
\vspace{-15pt}
\end{figure}

\begin{table*}[]
\small
\centering
\caption{Time complexity of different models on behavior prediction part.}
\vspace{-8pt}
\label{table:complexity}
\renewcommand{\arraystretch}{1.0}
 \setlength{\tabcolsep}{0.7mm}{ 
  \scalebox{0.8}{
\begin{tabular}{c|c|c}
\hline
Method         & Behavior Prediction                   & Remark                                                                                                                                                                                                                  \\ \hline
Bert4Rec       & $O(B*\rho*|\mathcal{T}|*K^{v})$       & $\rho$ is the ratio of sequence tokens for Cloze task                                                                                                                                                                   \\ \hline
CARCA          & $O(B*K^{v}+B*H'*(K^{v}*N')*d)$ & \begin{tabular}[c]{@{}c@{}}It involves cross attention between user-side features and candidate item.\\ $H'$ is number of heads in cross attention, \\ and $N'$ is the number of user-side features\end{tabular} \\ \hline
SceneCTC       & $O(B*K^{v})$                          & It has no contrastive loss                                                                                                                                                                                              \\ \hline
MSDP           & $O(B*K^{v}+B^2*d)$                    & It involves contrastive loss of input sequence                                                                                                                                                                          \\ \hline
ContraRec      & $O(B*K^{v}+B^2*d)$                    & It involves contrastive loss of input sequence                                                                                                                                                                          \\ \hline
SceneContraRec & $O(B*K^{v}+B^2*d)$                    & It involves contrastive loss of input sequence                                                                                                                                                                          \\ \hline
\rowcolor{improvementblue} DSPnet         & $O(B*K^{v}+2*B^2*d)$                  & It involves CCR loss of two input sequences                                                                                                                                                                             \\ \hline
\end{tabular}
}}
\vspace{-8pt}
\end{table*}
\section{Model Complexity Comparison}
\label{appen:complexity}
We analyze the complexity of different models via two components: feature encoding and behavior prediction, both of which are commonly present in sequential behavior prediction models. The compared methods all use the powerful and popular transformer encoder. 
Let $B$ be the batch size, $L$ be the number of transformer layers, $|\mathcal{T}|$ be the sequence length of samples, $H$ be the head number and $d$ be the dimension of each head. \textbf{The time complexity of transformer encoder can be represented as $O(B*L*H*|T|^2*d)$, which is nearly the same for all compared methods. The main difference of complexity lies in behavior prediction.} Before analyzing the complexity of behavior prediction part, we denote $K^{v}$ as the number of candidate items (including positive and negative ones) for prediction, the complexity comparison is listed in Table~\ref{table:complexity}.

In this table, $K^{v}$ generally reaches the magnitude of millions in industrial settings. The value of $|\mathcal{T}|$ varies depending on the dataset, and for our business applications, we typically set it to $100$. Consequently, the complexity of DSPnet is considerably lower than that of CARCA and Bert4Rec, and does not significantly increase over SOTA methods such as SceneContraRec. \textbf{This makes DSPnet well-suited for usage in large-scale industrial datasets. We have successfully deployed it in our system using 16 A100 GPUs.}

\section{More Experiments}
\label{sec:appendix_more_experiments}
\subsection{Period Item Prediction Performance}
The results of period item prediction task is shown in Table~\ref{table:period_item_prediction}.

\begin{table*}[]
\small
\centering
\caption{Performance comparison on period item prediction. R@$k$ and N@$k$ are Recall@$k$ and NDCG@$k$. We use ``w/o" to denote DSPnet without a particular part. The best results are bolded and \textit{the most competitive public baselines} are underlined.}
\vspace{-6pt}
\label{table:period_item_prediction}
\renewcommand{\arraystretch}{0.9}
 \setlength{\tabcolsep}{0.5mm}{ 
  \scalebox{0.85}{
\begin{tabular}{c|cccc|cccc|cccc}
\hline
Dataset               & \multicolumn{4}{c|}{Outbrain}                                                                                                                                                                                                                         & \multicolumn{4}{c|}{AllScenePay-1m} & \multicolumn{4}{c}{AllScenePay-10m} \\ \hline
Method                & R@5                                                         & N@5                                                         & R@10                                                        & N@10                                                        & R@5    & N@5    & R@10    & N@10    & R@5    & N@5    & R@10    & N@10    \\ \hline
BERT4Rec              & 0.0946                                                      & 0.0694                                                      & 0.1388                                                      & 0.084                                                       & {\small OOM}       & {\small OOM}       & {\small OOM}        & {\small OOM}        & {\small OOM}       & {\small OOM}       & {\small OOM}        & {\small OOM}        \\
MSDP                                                                           & 0.2671                                                           & 0.2191                                                          & 0.2967                                                          & 0.229                                                           & 0.0007                                                         & 0.0006                                                         & 0.0011                                                         & 0.0008                                                         & 0.0008  & 0.0007  & 0.0012 & 0.0009 \\
ContraRec                                                                      & 0.3573                                                           & 0.2481                                                          & 0.4654                                                          & 0.2844                                                          & 0.0759                                                         & 0.0790                                                         & 0.0888                                                         & 0.0839                                                         &  0.1534       & \underline{0.1610}        & 0.1796       &  \underline{0.1707}      \\
SceneCTC                                                                       & 0.4754                                                           & 0.4080                                                 & 0.5184                                                          & 0.4225                                                & 0.0758                                                         & 0.0785                                                         & 0.0913                                                         & 0.0848                                                         & \underline{0.1538}  & 0.1585  & 0.1813 & 0.1694 \\
SceneContraRec                                                                 & 0.4902                                                 & 0.4031                                                          & 0.5388                                                & 0.4197                                                          & \underline{0.0790}                                               & \underline{0.0821}                                               & \underline{0.0933}                                               & \underline{0.0876}                                               & 0.1532  & 0.1587  & \underline{0.1820}  & 0.1702 \\ 
CARCA                                                                           & \underline{0.5092}                                                           & \underline{0.4393}                                                          & \underline{0.5402}                                                         & \underline{0.4499}                                                           & {\small OOM}                                                         & {\small OOM}                                                         & {\small OOM}                                                         & {\small OOM}                                                         & {\small OOM}  & {\small OOM}  & {\small OOM} & {\small OOM} \\
\hline
DSPnet--                                                                 & 0.5281                                             & 0.4634                                                           & 0.5570                                                 & 0.4372                                                           & 0.0867                                                & 0.0905                                                & 0.1029                                                & 0.0968                                                & 0.1636 & 0.1703  & 0.1922 & 0.1812 \\
DSPnet({\small w/o $\mathcal{L}_{\text{APR}}$, $\mathcal{L}_{\text{CCR}}$}) & 0.6062                                                           & 0.5307                                                          & 0.6583                                                          & 0.5482                                                          & 0.0997                                                         & 0.1053                                                         & 0.1121                                                         & 0.1090                                                         & 0.1895  & 0.1998  & 0.2149 & 0.2083 \\
DSPnet({\small w/o $\mathcal{L}_{\text{CCR}}$})                                & 0.6069                                                           & 0.5347                                                          & 0.6635                                                          & 0.5537                                                          & 0.0995                                                         & 0.1049                                                         & 0.1120                                                         & 0.1088                                                         & 0.1930   & 0.2033  & 0.2186 & \textbf{0.2118} \\
DSPnet({\small w/o $\mathcal{L}_{\text{APR}}$})                                & 0.6151                                                           & \textbf{0.5396}                                                          & 0.6651                                                          & \textbf{0.5562}                                                          & 0.1007                                                         & 0.1060                                                         & 0.1148                                                         & 0.1107                                                         & \textbf{0.1930}   & \textbf{0.2037}  & 0.2178 & 0.2117 \\ 
\hline
\rowcolor{improvementblue} DSPnet                   & \begin{tabular}[c]{@{}c@{}}\textbf{0.6198}\\ {\small (+11.06\%)}\end{tabular} & \begin{tabular}[c]{@{}c@{}}0.5388\\ {\small(+9.95\%)}\end{tabular} & \begin{tabular}[c]{@{}c@{}}\textbf{0.6682}\\ {\small(+12.80\%)}\end{tabular} & \multicolumn{1}{c|}{\begin{tabular}[c]{@{}c@{}}0.5549\\ {\small(+10.50\%)}\end{tabular}} &  \begin{tabular}[c]{@{}c@{}}\textbf{0.1015}\\ {\small(+2.25\%)}\end{tabular} & \begin{tabular}[c]{@{}c@{}}\textbf{0.1071}\\ {\small(+2.50\%)}\end{tabular} & \begin{tabular}[c]{@{}c@{}}\textbf{0.1149}\\ {\small(+2.16\%)}\end{tabular} & \multicolumn{1}{c|}{\begin{tabular}[c]{@{}c@{}}\textbf{0.1114}\\ {\small(+2.38\%)}\end{tabular}}        &   \begin{tabular}[c]{@{}c@{}}0.1926\\ {\small(+3.88\%)}\end{tabular} & \begin{tabular}[c]{@{}c@{}}0.2028\\ {\small(+4.18\%)}\end{tabular} & \begin{tabular}[c]{@{}c@{}}\textbf{0.2187}\\ {\small(+3.67\%)}\end{tabular} & \multicolumn{1}{c}{\begin{tabular}[c]{@{}c@{}}0.2115\\ {\small(+4.08\%)}\end{tabular}}       \\ \hline
\end{tabular}
}}
\vspace{-8pt}
\end{table*}

\subsection{Hyper-Parameter Sensitivity}
In our DSPnet, $\alpha$ and $\beta$ control the weight on adversarial prior regularization and conditional contrastive regularization, respectively. We here investigate the effects on performance of these two hyper-parameters. The results are collected in Figure~\ref{figure:hyper_params}.

From this figure, when comparing different columns, we see that the model performance varies a lot, highlighting the substantial impact of the loss weight on CCR. Additionally, the optimal hyper-parameter settings differ, primarily due to the distinct data distributions of Outbrain and AllScenePay-1m. It is practical to set these hyper-parameters according to the used data.

\begin{table*}[]
\centering
\caption{Study of the prior distribution.} 
\vspace{-8pt}
\label{table:prior_distribution}
\renewcommand{\arraystretch}{0.9}
 \setlength{\tabcolsep}{0.7mm}{ 
  \scalebox{0.9}{
\begin{tabular}{c|cccc|cccc}
\hline
Dataset           & \multicolumn{4}{c|}{Ourbrain}                                        & \multicolumn{4}{c}{AllScenePay-1m}    \\ \hline
Model             & R@5             & N@5             & R@10            & N@10           & R@5             & N@5             & R@10            & N@10    \\ \hline
Uniform           & 0.6025          & 0.5190           & 0.6537          & 0.5355         & 0.0851          & 0.0621          & 0.1129          & 0.0711           \\
Laplace           & 0.5867          & 0.5031          & 0.64            & 0.5206         & 0.0839          & 0.0610           & 0.1127          & 0.0703           \\
Multi-Gaussian         & 0.5982          & 0.5181          & 0.6485          & 0.5343         & 0.0842          & 0.0618          & 0.1114          & 0.0705            \\
Lognormal         & 0.6177          & 0.5373          & 0.6648          & 0.5525         & 0.0843          & 0.0615          & 0.1131          & 0.0708           \\ \hline
\rowcolor{improvementblue} Standard Guassian & \textbf{0.6248} & \textbf{0.5368} & \textbf{0.6717} & \textbf{0.5520} & \textbf{0.0870} & \textbf{0.0632} & \textbf{0.1155} & \textbf{0.0725}     \\ \hline
\end{tabular}
}}
\vspace{-8pt}
\end{table*}
\subsection{Effects of Different Priors}
\label{sec:appendix_prior}
DSPnet incorporates adversarial prior regularization loss to impose prior knowledge on the learned representations. We explore the influence of different prior distributions on model performance. In particular, we use the uniform distribution $\mathcal{U}(0,1)$, the Laplace distribution $\text{Laplace}(0,1)$, the standard normal distribution $\mathcal{N}(0,1)$, and the lognormal distribution $\text{Lognormal}(0,1)$. Additionally, we used a sum of two normal distributions, $\mathcal{N}(0,1) + \mathcal{N}(3,1)$, to construct Multi-Gaussian distribution.
The results are shown in Table~\ref{table:prior_distribution}.

By analyzing the results from this table, we see that the standard Gaussian distribution consistently shows the best performance on three datasets. This observation matches a widely accepted principle in recommendation and search systems, where user preferences tend to exhibit Gaussian distribution~\cite{10.1145/3178876.3186150,cui2018variational,Xie2021}. A multi-Gaussian approach has the potential to capture user preferences more accurately, because individuals often have multiple areas of interest. However, determining the parameters for a Multi-Gaussian model can be rather complex. Consequently, employing the standard Gaussian serves as a easy and effective choice in practice.

\begin{table*}[]
\centering
\caption{Performance comparison on next scene prediction task. R@$k$ and N@$k$ represent Recall@$k$ and NDCG@$k$, respectively.}
\vspace{-8pt}
\label{table:next_scene_prediction}
\renewcommand{\arraystretch}{0.9}
 \setlength{\tabcolsep}{0.8mm}{ 
  \scalebox{0.95}{
\begin{tabular}{ccccc|cccc}
\hline
Dataset        & \multicolumn{4}{c|}{Outbrain} & \multicolumn{4}{c}{AllScenePay-1m} \\ \hline
Method         & R@5   & N@5   & R@10  & N@10  & R@5    & N@5    & R@10    & N@10   \\ \hline
BERT4REC       & 0.3043     & 0.2763    & 0.3345     & 0.2861   & 0.8253  & 0.6365     & 0.9205     & 0.6483    \\
MSDP           & 0.2638     & 0.1798     & 0.3789     & 0.2167     & 0.8464      & 0.6489     & 0.9371      & 0.6788     \\
ContraRec      & 0.4071     & 0.3468     & 0.4662     & 0.3661     & 0.8710      & 0.6635      & 0.9508       & 0.6903      \\
SceneCTC       & 0.6246     & 0.5492     & 0.6937     & 0.5715     & 0.8872      & 0.6692      & 0.9547       & 0.6918      \\
SceneContraRec & 0.6175     & 0.5386     & 0.6858     & 0.5606     & 0.8692     & 0.6629      & 0.9528      & 0.6911     \\ \hline
\rowcolor{improvementblue} DSPnet          & \textbf{0.6567}    & \textbf{0.5770}     & \textbf{0.7200}     & \textbf{0.5975}     &  \textbf{0.8944}    & \textbf{0.6816}      & \textbf{0.9629}       & \textbf{0.7045}      \\ \hline
\end{tabular}
}}
\end{table*}
\subsection{On the Prediction Performance of Scene}
We also conducted additional experiments to evaluate the scene prediction capabilities of various models, aiming to demonstrate that our DSPnet more effectively captures the ``scene" aspect alongside item sequences. The results, presented in Table~\ref{table:next_scene_prediction}, indicate that DSPnet achieves better scene prediction ability, validating the effectiveness of our model.

\begin{table}[]
\centering
\caption{Results of combining CCR with other models.}
\vspace{-8pt}
\label{table:combineCCR}
\renewcommand{\arraystretch}{1.2}
 \setlength{\tabcolsep}{0.7mm}{ 
  \scalebox{0.75}{
\begin{tabular}{ccccc|cccc}
\hline
Dataset        & \multicolumn{4}{c|}{Outbrain} & \multicolumn{4}{c}{AllScenePay-1m} \\ \hline
Method         & R@5   & N@5   & R@10  & N@10  & R@5    & N@5    & R@10    & N@10   \\ \hline
SceneContraRec & \textbf{0.4979}     & 0.4027     & 0.5448     & 0.4182     & 0.0762     & 0.0544      & 0.1045      & 0.0635     \\ \hline
\rowcolor{improvementblue} SceneContraRec+CCR          & 0.4966    & \textbf{0.4088}     & \textbf{0.5587}     & \textbf{0.4250}     &  \textbf{0.0787}    & \textbf{0.0571}      & \textbf{0.1033}       & \textbf{0.0651}      \\ \hline
\end{tabular}
}}
\end{table}
\subsection{Combining CCR with Other Models}
\label{append:appen_combineCCR}
In order to more easily show the advantages of CCR loss over conventional contrastive loss, we replaced the contrastive regularization in SceneContraRec with CCR loss. The comparison results are summarized in Table~\ref{table:combineCCR}. In this table, the only difference between ``SceneContraRec'' and ``SceneContraRec+CCR'' is that ``SceneContraRec+CCR'' utilize our CCR loss to learn representation invariance, while ``SceneContraRec'' employs the conventional contrastive loss. The comparison results can well demonstrate that CCR does more benefits to the model learning. The learnable conditional weights in CCR allow better representation invariance learning, especially on skewed user behavior distributions.

\end{document}